\documentclass[12pt]{report}
\usepackage{amsmath,amssymb,cite}
\usepackage{graphicx}
\usepackage{latexsym}
\usepackage{dcolumn}
\usepackage{enumitem} 
\usepackage{bm}
\title{Cosmology In Terms Of The Deceleration Parameter.  Part II}
\author{Yu.L. Bolotin, V.A. Cherkaskiy, O.A. Lemets \\ D.A. Yerokhin and L.G. Zazunov}
\date{\flushright{\small''All of observational cosmology is the search\\ for two numbers: $H_0$ and $q_0$.''\\ Allan Sandage, 1970}}
\begin{document}
\maketitle
\begin{abstract}

In the early seventies, Alan Sandage defined cosmology as the search for two
numbers: Hubble parameter ${{H}_{0}}$ and deceleration parameter ${{q}_{0}}$.
The first of the two basic cosmological parameters (the Hubble parameter)
describes the linear part of the time dependence of the scale factor. Treating
the Universe as a dynamical system it is natural to assume that it is
non-linear: indeed, linearity is nothing more than approximation, while
non-linearity represents the generic case. It is evident that future models of
the Universe must take into account different aspects of its evolution. As soon
as the scale factor is the only dynamical variable, the quantities which
determine its time dependence must be essentially present in all aspects of the
Universe' evolution. Basic characteristics of the cosmological evolution, both
static and dynamical, can be expressed in terms of the parameters ${{H}_{0}}$
and ${{q}_{0}}$. The very parameters (and higher time derivatives of the scale
factor) enable us to construct model-independent kinematics of the cosmological
expansion.

Time dependence of the scale factor reflects main events in history of the
Universe. Moreover it is the deceleration parameter who dictates the expansion
rate of the Hubble sphere and determines the dynamics of the observable galaxy
number variation: depending on the sign of the deceleration parameter this
number either grows (in the case of decelerated expansion), or we are going to
stay absolutely alone in the cosmos (if the expansion is accelerated).

The intended purpose of the report is reflected in its title --- "Cosmology
in terms of the deceleration parameter". We would like to show that practically
any aspect of the cosmological evolution is tightly bound to the deceleration
parameter.

It is the second part of the report. The first part see here http://arxiv.org/abs/1502.00811
\end{abstract}
\tableofcontents
\newpage
\chapter{Deceleration Parameter in Different Cosmological Models}
\section{SCM}
We now consider the evolution of the deceleration parameter (DP) in the standard cosmological model (SCM) \cite{padmanabhan,sahni,tkachev,copeland_sami_tsujikava,gorbunov_rubakov}.We recall that in the Big Bang model, the substances that filled up the Universe - matter and radiation - could only provide decelerating expansion. According to the SCM, presently the Universe is dominated by the dark energy (DE) -- the component with negative pressure. It is that component which leads to the observable accelerated expansion. Let us determine the redshift and transition time to the accelerating expansion, i.e., find the inflection point in the curve describing the time dependence of the scale factor. In the SCM, the second Friedmann equation
\[\frac{\ddot a}{a}=-\frac{1}{6M^2_{Pl}}(\rho+3p)\] can be reduced to the form
\begin{equation}\label{SCM_15_1}
\frac{\ddot a}{a}=\frac12H_0^2[2\Omega_{\Lambda0}-\Omega_{m0}(1+z)^3]
\end{equation}
where $\Omega_{\Lambda0}$ and $\Omega_{m0}$ are the present-day values of the relative DE density in the form of a  cosmological constant and of the matter density. Hence, we obtain the redshift value $z_t$ at which the transition from decelerating to accelerating expansion took place as
\begin{equation}\label{SCM_15_2}
z_t=\left(\frac{2\Omega_{\Lambda0}}{\Omega_{m0}}\right)^{1/3}-1.
\end{equation}
For the SCM parameters $\Omega_{\Lambda0}\simeq0.73$ and $\Omega_{m0}\simeq0.27$, we have $z_t\simeq0.745$. We note that result (\ref{SCM_15_2}) can be obtained using the fact that for two-component SCM Universe the DP is
\begin{equation}\label{SCM_15_3}
q=\frac12 -\frac32\frac{\Omega_{\Lambda0}}{(1+z)^3\Omega_{m0}+\Omega_{\Lambda0}}.
\end{equation}
The condition $q=0$ allows to reproduce (\ref{SCM_15_2}).

Let us find asymptotes of the DP in the SCM. For early Universe ($z\to\infty$) filled with positive pressure components $q(z\to\infty)=1/2$, i.e. the expansion is decelerated as was expected, while in far future with domination of the cosmological constant one observes the accelerated expansion $q(z\to-1)=-1$. The latter result is a trivial consequence of the exponential expansion of Universe $a(t)\propto e^{Ht}$ in the case of domination of DE in form of the cosmological constant.

\begin{figure}
\includegraphics[width=\textwidth]{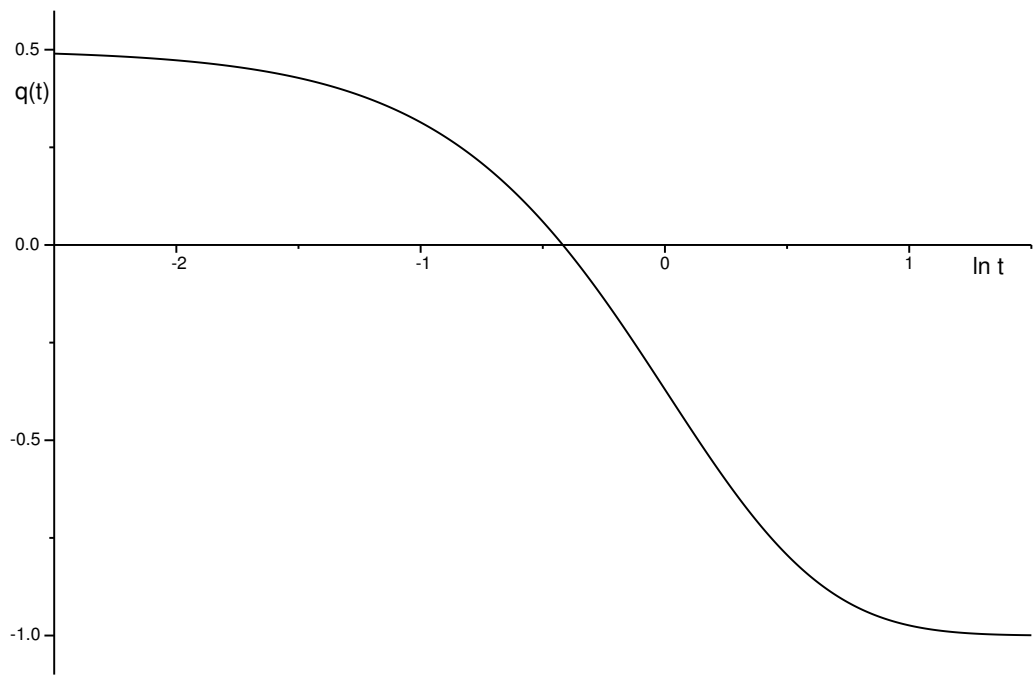}
\caption{\label{SCM_f1} Time dependence of the DP in the SCM.}
\end{figure}

Note that the dependence $q(t)$ can be obtained immediately from the definition $q\equiv-\ddot a/(aH^2)$ making use of the SCM solutions for the scale factor
\begin{align}\label{SCM_15_4}
a(t) & =A^{1/3}\sinh^{2/3}(t/t_\Lambda);\\
\nonumber
A\equiv & \frac{\Omega_{m0}}{\Omega_{\Lambda0}},\quad t_\Lambda\equiv\frac23H_0^{-1}\Omega_{\Lambda0}^{-1/2}.
\end{align}
As the result one obtains
\begin{equation}\label{SCM_15_5}
q(t)=\frac12\left[1-3\tanh^2\left(\frac t{t_\Lambda}\right)\right].
\end{equation}
The dependence $q(t)$ is presented on Fig.\ref{SCM_f1}

Let us now define the time corresponding to the transition to accelerated expansion. Inverting (\ref{SCM_15_4}) one obtains
\begin{equation}\label{SCM_15_6}
t(a)=\frac23\Omega_{\Lambda0}^{-1/2}H_0^{-1}{\rm arsh}\left[\left(\frac{\Omega_{\Lambda0}}{\Omega_{m0}}\right)^{1/2}a^{3/2}\right].
\end{equation}
Transforming from the redshift to the scale factor \[a^*=(1+z^*)^{-1}\left(\frac{\Omega_{m0}}{2\Omega_{\Lambda0}}\right)^{1/3}\] one finds
\begin{equation}\label{SCM_15_7}
t^*\equiv t(a^*)=\frac23\Omega_{\Lambda0}^{-1/2}H_0^{-1}{\sinh^{-1}} (1/2)\simeq5.25\ Gyr.
\end{equation}
Because of physical importance of the obtained result let us give one more (seemingly the simplest) its interpretation. If the quantity $aH=\dot a$ grows then $\ddot a>0$ which corresponds to accelerated expansion of the Universe. According to the first Friedmann equation
\begin{equation}\label{SCM_15_8}
\frac{aH}{H_0}=\sqrt{\frac{a^3\Omega_{\Lambda0}+\Omega_{m0}}{a}}\simeq\sqrt{\frac{0.73a^3+0.27}{a}}.
\end{equation}
It is easy to show that the function above starts to grow at $a^*=0.573$, which corresponds to $z^*=0.745$. It is worth noting that the transition to accelerates expansion of the Universe ($z^*=0.75$) occurred considerably earlier then the transition to the DE domination epoch ($z^*=0.4$).

A natural question arises: what is limiting value of the equation of state (EoS) parameter $w$ which still provides accelerated expansion of the Universe at present time. As we have seen above, the condition of the accelerated expansion reads \[\sum\limits_i(\rho_i+3p_i)<0.\] In the SCM this condition transforms to
\begin{equation}\label{SCM_15_9}
w_{de}<-\frac13\Omega_{de}^{-1};\quad w_{de}<0.46.
\end{equation}
Of course the substance with such EoS is distinct from the cosmological constant and van be realized for example with the help of the scalar fields.

Let us now use the SCM parameters to estimate the absolute value of the cosmological acceleration. Taking derivative of the Hubble law w.r.t. time one obtains
\begin{equation}\label{SCM_15_10}
\dot V=(\dot H+H^2)R.
\end{equation}
Time derivative of the Hubble parameter
\begin{equation}\label{SCM_15_11}
\dot H=\frac{\ddot a a-\dot a^2}{a^2}=\frac{\ddot a}{a}-H^2.
\end{equation}
Then
\begin{equation}\label{SCM_15_12}
\dot H +H^2=\frac{\ddot a}{a}=-\frac1{6M_{Pl}^2}(\rho_m-2\rho_\Lambda) = \frac1{3M_{Pl}^2}\left(\rho_\Lambda -\frac12\rho_m\right)=H^2\left(\Omega_\Lambda -\frac12\Omega_m\right).
\end{equation}
Finally one gets for the acceleration $\dot V$ an analogue of the Hubble law
\begin{equation}\label{SCM_15_13}
\dot V=\tilde H R;\quad \tilde H=H^2\left(\Omega_\Lambda -\frac12\Omega_m\right).
\end{equation}
For instance, at distance $R=1\ Mpc$ one finds $\dot V\simeq10^{-11}cm/sec^2$.

\begin{figure}
\includegraphics[width=\textwidth]{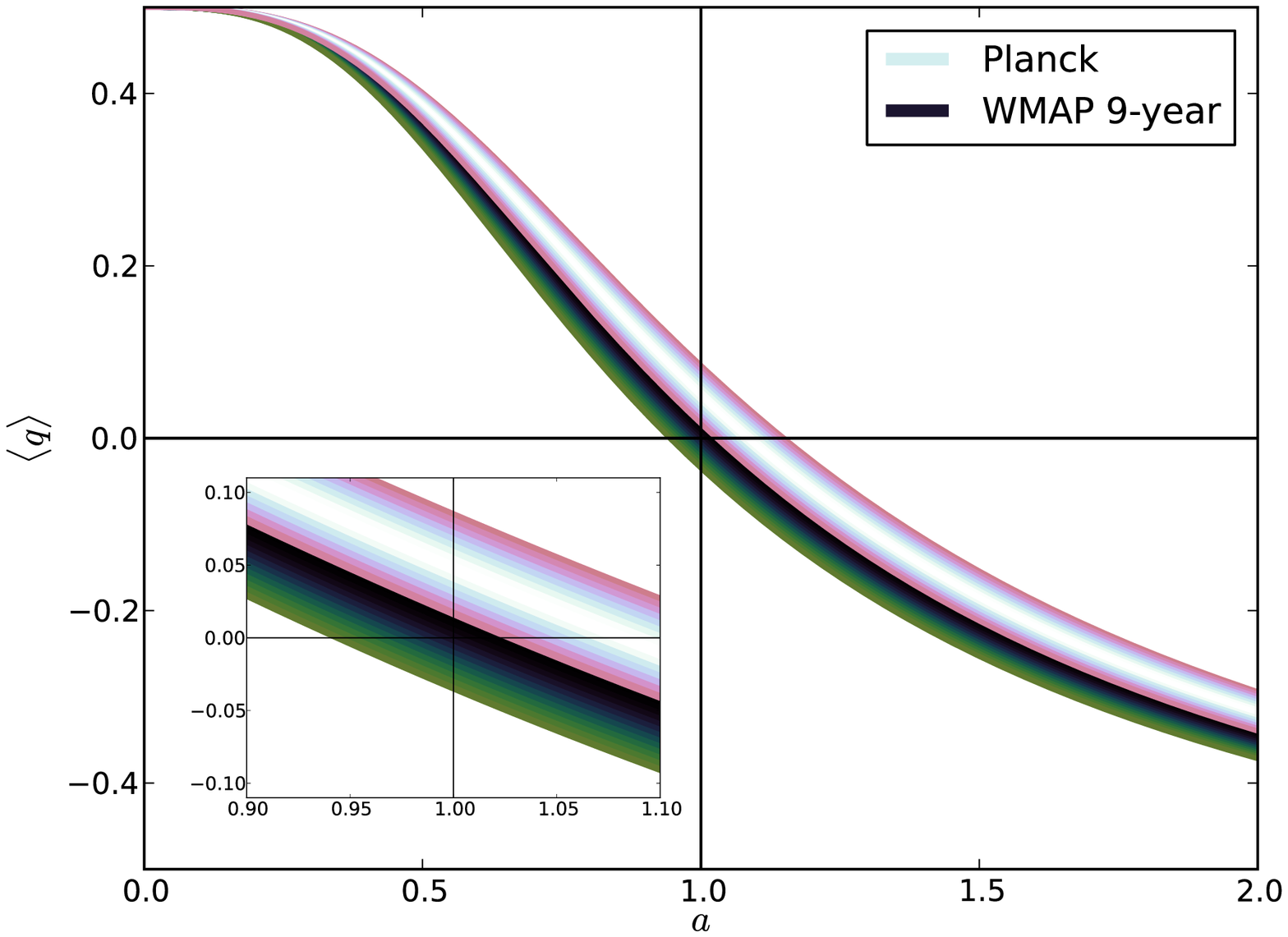}
\caption{\label{SCM_f2} Comparing $\bar q$ as a function of the cosmological scale factor, $a(t)$, using the WMAP $9$-year and Planck data \cite{van_oirschot_kwan_lewis}.}
\end{figure}

Let us now turn back to the hypothesis about equality to zero of the average DP (see Part I, section 2.5), expressed by Lima \cite{lima}. Over the last nine years of cosmic microwave background observations, the Wilkinson Microwave Anisotropy Probe (WMAP) results were consistent with the $\Lambda$CDM cosmological model in which the age of the Universe is one Hubble time, and the time-averaged value of the DP is consistent with zero. As was noted in \cite{bolotin_erokhin_lemets}, this curious observation has been put forward as a new coincidence problem for the $\Lambda$CDM concordance cosmology, which is in fact a 'greater' coincidence than the near equality of the density parameters of matter and the cosmological constant. However recent Planck's results \cite{planck_2013_results} make the new coincidence problem not so important. Let us convince ourself in this. Setting $t=t_0$ and $a=a_0=1$ in (\ref{SCM_15_6}), the $H_0t_0=1$ condition transforms to
\begin{equation}\label{SCM_15_14}
e^{-3\sqrt{\Omega_{\Lambda0}}}(\Omega_{\Lambda0}-1) + e^{3\sqrt{\Omega_{\Lambda0}}}(\Omega_{\Lambda0}-1) + 2\Omega_{\Lambda0}+2=0.
\end{equation}
The only positive solution to equation (\ref{SCM_15_14}) is $\Omega_{\Lambda0}\approx0.737125$. In general, we can write the time-averaged DP $\langle q\rangle$ as a function of scale factor:
\begin{equation}\label{SCM_15_15}
\langle q\rangle+1=\frac1{tH}=\left[\frac23\sqrt{\frac{\Omega_{m0}}{\Omega_{\Lambda0}a^3}+1}\cosh^{-1}\sqrt{1+ \frac{a^3\Omega_{\Lambda0}}{\Omega_{m0}}}\right]^{-1}.
\end{equation}
Figure \ref{SCM_f2} shows the $95\%$ confidence levels of $\Omega_{\Lambda0}$ from both the Planck data ($0.686^{+0.037}_{-0.040}$) and the WMAP $9$-year data ($0.721\pm0.050$). The probability density function of the posterior distribution is indicated by the colorscale. Note that $\bar q(t_0)$ is close to zero only during a brief period in cosmic time. Moreover, $\bar q(t_0)\approx0$ in the WMAP 9-year data,  however, $\bar q(t_0)\ne0$ for the Planck data.
\section{Cosmology with power and hybrid expansion laws}
Let us consider a general class of power-law cosmology \cite{kumar,rani,dolgova_halenkad_tkachev} described by the scale factor
\begin{equation}\label{SCM_16_1}
a(t)=a_0\left(\frac t{t_0}\right)^\alpha,
\end{equation}
where $\alpha$ is a dimensionless positive parameter. The DE in this case
\begin{equation}\label{SCM_16_2}
q\equiv-\frac{\ddot a}{aH^2}=\frac1\alpha-1.
\end{equation}
The positivity of $\alpha$ leads to $q>-1$. The scale factor in terms of the DP is
\begin{equation}\label{SCM_16_3}
a(t)=a_0\left(\frac t{t_0}\right)^{\frac1{q+1}},
\end{equation}
We observe that $q>-1$ is the condition for expanding Universe in the power-law cosmological model. The expansion history of the Universe in power-law cosmology depends on the parameters $H_0$ and $q_0$
\begin{equation}\label{SCM_16_4}
H(z)=H_0(1+z)^{1+q}.
\end{equation}
The comoving distance (1406.2445)
\begin{align}\nonumber
r(z) & =\frac1{\sqrt{|\Omega_c|}}F\left(|\Omega_c|\int\limits_0^zdz'\frac{H_0}{H(z')}\right);\\
\label{SCM_16_5}\Omega_c & \equiv-\frac{k}{a_0^2H_0^2};\\ \nonumber
F(x) & \equiv\left\{
\begin{array}{cc}
\sinh(x), & k=-1\\
x, & k=0\\
\sin(x), & k=1
\end{array}
\right.
\end{align}
with substitution (\ref{SCM_16_5}) can be transformed to
\begin{equation}\label{SCM_16_6}
r(z) =\frac1{\sqrt{|\Omega_c|}}F\left(|\Omega_c|q^{-1}[1-(1+z)^{-q}]\right).
\end{equation}
Note that the Milne model can be viewed as a power-law cosmology with $\Omega_c=1$, $\alpha=1$ or $q=0$. For this model therefore
\begin{equation}\label{SCM_16_7}
r(z) =\sinh[\ln(1+z)].
\end{equation}
In power-law cosmology the scale factor $a(t)$ and the CMB temperature $T(t)$ are related through the relation
\begin{equation}\label{SCM_16_8}
\frac{T_0}T=\frac a{a_0}=\left(\frac t{t_0}\right)^\alpha=\left(\frac t{t_0}\right)^{\frac1{1+q}}.
\end{equation}
Consequently,
\begin{equation}\label{SCM_16_9}
T(t)=T_0\left(\frac t{t_0}\right)^{-\alpha}=T_0\left(\frac t{t_0}\right)^{-\frac1{1+q}}.
\end{equation}
Let us consider now a simple generalization of power-law cosmology, called the hybrid expansion law \cite{akarsu}
\begin{equation}\label{SCM_16_10}
a(t)=a_0\left(\frac t{t_0}\right)^\alpha e^{\beta\left(\frac t{t_0}-1\right)},
\end{equation}
where $\alpha$ and $\beta$ are non-negative constants. Further, $a_0$ and $t_0$ as well as in the power-law cosmology respectively denote the scale factor and age of the Universe today. The cosmographical parameters : Hubble parameter, DP  and jerk parameter are respectively
\begin{align}\nonumber
H & =\frac\alpha t+\frac\beta{t_0};\\
\label{SCM_16_11} q & =\frac{\alpha t^2_0}{(\beta t+\alpha t_0)^2}-1;\\ \nonumber
j & = \frac{\alpha t^2_0(2t_0-3\beta t-3\alpha t_0)}{(\beta t+\alpha t_0)^3}+1.
\end{align}
In particular cases, one obviously obtains power-law and exponential expansion from (\ref{SCM_16_10}) choosing $\alpha=0$ and $\beta=0$ respectively. Accordingly, for $t\to0$, the cosmological parameters approximate to the following:
\begin{equation}\label{SCM_16_12}
a\to a_0\left(\frac t{t_0}\right)^\alpha,\quad H\to\frac\alpha t,\quad q\to -1+\frac1\alpha,\quad j \to 1-\frac3\alpha+\frac2{\alpha^2}.
\end{equation}
Similarly, the exponential term dominates at late times, such that in the limit $t\to\infty$ we have
\begin{equation}\label{SCM_16_13}
a\to a_0e^{\beta\left(\frac t{t_0}-1\right)},\quad H\to\frac\beta{t_0},\quad q\to-1,\quad j\to1.
\end{equation}
It may be observed that the parameter $\alpha$ determines the initial kinematics of the Universe while the very late time kinematics of the Universe is determined by the parameter $\beta$. When $\alpha$ and $\beta$ both are non-zero, Universe evolves with variable DP given by (\ref{SCM_16_11}) and the transition from deceleration to acceleration takes place at \[\frac t{t_0}\simeq\frac{\sqrt\alpha-\alpha}{\beta},\] which puts $\alpha$ in the range $0<\alpha<1$.
\section{Cosmological models with constant deceleration parameter (Berman model)}
In 1983, Berman \cite{berman} proposed a special law of variation of Hubble parameter in FLRW space-time, which yields a constant value of DP. Such a law of variation for Hubble's parameter is not inconsistent with the observations and is also approximately valid for slowly time-varying DP models. The law provides explicit forms of scale factors governing the FLRW Universe and facilitates to describe accelerating as well as decelerating modes of evolution of the Universe. Models with constant DP have been extensively studied in the literature in different contexts \cite{kumar2,akarsu1,akarsu2,akarsu3,yadav,yadav2}.

Let us consider a law of variation for the Hubble parameter proposed by Berman, which yields a constant value of DP,
\begin{equation}\label{SCM_17_1}
H=Da^{-n}.
\end{equation}
Integration of this relation leads to
\begin{equation}\label{SCM_17_2}
a(t)=(nDt+c_1)^{\frac1n}.
\end{equation}
where $c_1$ is a constant of integration. The value of DP in this case is
\begin{equation}\label{SCM_17_3}
q=n-1
\end{equation}
which is a constant. As usual, sign of $q$ determines the expansion regime. A positive sign of $q$, i.e., $n>1$ corresponds to the standard decelerating model whereas the negative sign of $q$, i.e., $0<n<1$ corresponds to accelerating model. The expansion of the Universe at a constant rate corresponds to $n=1$, that is, $q=0$. Therefore in the considered model one can choose the values of DP consistent with the observations.
\section{Linearly varying deceleration parameter}
We have considered above the simplest case --- the time-independent DP. Such situation clearly contradicts to our understanding of the Universe's evolution: changes of epochs must inevitably lead to variation of the DP. Recall that in the standard $\Lambda$CDM model, the DP is variable and evolves from $1/2$ to $-1$. Linear parametrization of the DP represents quite naturally the next logical step. In particular, the parametrizations $q=q_0+q_1z$ and $q=q_0+q_1(1-a_0/a)$ of DP, which are linear in cosmic redshift $z$ and scale factor $a$, have been frequently utilized in the literature \cite{riess,cunha_lima,xu_liu,xu_lu,cunha}.

A more general approach is to expand the DP in Taylor series
\begin{equation}\label{SCM_18_1}
q(x)=q_0+q_1\left(1-\frac x{x_0}\right)+q_2\left(1-\frac x{x_0}\right)^2+\ldots
\end{equation}
where $x$ is a some cosmological parameter as cosmic scale factor $a$, cosmic redshift $z$, cosmic time etc. The next step is to choose the parameter to decompose in and number of the terms kept. First of all the choice is dictated by the adequate description of the observational data, which of course will improve when the number of terms in Taylor expansion is increased. However increasing the number of terms will lead to growth of the model parameter number and make it less efficient. Decreasing of the decomposition terms number can be possibly compensated by a lucky choice of the cosmological parameter to decompose in. Therefore as the first step in utilization of the decomposition (\ref{SCM_18_1}) (it is natural to call this procedure the parametrization of the DP) one can take the following linear approximation
\begin{equation}\label{SCM_18_2}
q(x)=q_0+q_1\left(1-\frac x{x_0}\right)
\end{equation}
Detailed analysis of the linear approximation (\ref{SCM_18_1}) was performed in \cite{akarsu_dereli,akarsu_dereli_kumar_xu}.

Let us start from consideration of the linearly varying DP in terms of cosmic redshift $z$.  In this case \[\frac x{x_0}=\frac{a_0}a=1+z\] and
\begin{equation}\label{SCM_18_3}
q=q_0+q_1z.
\end{equation}

We note that the DP grows monotonically with no limits as we go back to earlier times of the Universe. On the other hand, the model is well behaved in the future such that the Universe either expands forever (provided that $q_0\ge q_1-1$) or ends with a Big Rip in finite future (provided that $q_0<q_1-1$). However, the predicted future of the Universe using (\ref{SCM_18_3}) cannot be considered so reliable since the observational constraints obtained using it with high redshift data are not reliable.

Unlimited growth of the DP for large $z$ in the parametrization (\ref{SCM_18_3}) forces us to consider the  linearly varying DP  in terms of  scale factor which can be obtained by substituting \[\frac x{x_0}=\frac a{a_0},\]
\begin{equation}\label{SCM_18_4}
q=q_0+q_1\left(1-\frac a{a_0}\right).
\end{equation}
Such parametrization can be written in terms of redshift as follows:
\begin{equation}\label{SCM_18_5}
q=q_0+q_1\frac z{1+z}.
\end{equation}
Treating the Universe as a dynamical system it is useful, along with the parametrizations (\ref{SCM_18_3})-(\ref{SCM_18_5}), to consider the parametrization of the DP directly in terms of cosmic time $t$. Such parametrization can be obtained by substitution \[\frac x{x_0}=\frac t{t_0}\] in (\ref{SCM_18_2}):
\begin{equation}\label{SCM_18_6}
q=q_0+q_1\left(1-\frac t{t_0}\right).
\end{equation}
In contrast to the two previous models, this model cannot be used for observational analysis directly. One should first solve it for the scale factor explicitly, and then obtain the time red-shift relation. This procedure can be facilitated if one first solve the equation with separating variables
\begin{equation}\label{SCM_18_7}
\frac{dH}{dt}=-H^2(1+q).
\end{equation}
Substituting (\ref{SCM_18_6}), one finds
\begin{equation}\label{SCM_18_8}
H(t)=2\frac{t_0}{t}[2(1+q_0+q_1)t_0-q_1t]^{-1}.
\end{equation}
Using the obtained expression for $H(t)$ to integrate the equation
\begin{equation}\label{SCM_18_9}
\frac{dz}{dt}=-(1+z)H(t)
\end{equation}
we obtain:
\begin{equation}\label{SCM_18_10}
q=q_0+q_1\left(1-\frac{2(1+q_0+q_1)}{q_1+(2+2q_0+q_1)(1+z)^{1+q_0+q_1}}\right).
\end{equation}

There are many ways to parametrize the DP with different forms of EoS for components filling the Universe. Let us consider for instance a single-component flat Universe with a fluid described by an EoS parameter expressed as a first order Taylor expansion in cosmic time \cite{akarsu_dereli_vazquez}, that is:
\[w=w_0+w_1(1-t),\]
where $w_0$ and $w_1$ are real constants and $t$ is the normalized time. This parametrization in the single-component flat ($\Omega=1$) case gives for the DP
\[q=\frac\Omega2+\frac32\sum\limits_iw_i\Omega_i=\frac12+\frac32[w_0+w_1(1-t)].\]
Such an approach enables us to use the cosmographic (model-free) constraints to test the EoS for different components.
\chapter{Kinematic Aspects of Dynamical Forms of Dark Energy}
\section{Quintessence}
The cosmological constant is just one of possible realizations of the hypothetical substance --- the DE, introduced to explain the accelerated expansion of the Universe. As we have seen above, such a substance must have the EoS $p=w\rho$ with parameter $w$ which satisfies the condition $w<-1/3$ (in absence of other components). Unfortunately the nature of the DE is mysterious for us and that is why there is huge number of hypotheses and candidates for this role of the fundamental component of energy budget of the Universe. We mentioned many times about the fast progress of observational cosmology during the last century. However we still cannot answer the question of time evolution of the parameter $w$. If this parameter does depend on time then we have to look for an alternative to the cosmological constant. For a short time a great number of the alternative possibilities (w.r.t $w=-1$) was studied. Scalar fields which formed the post-inflationary Universe are one of the main candidates for role of the DE. A version of scalar field $\varphi$ with appropriately chosen potential $V(\varphi)$ is the most popular one. In such models, in distinction to the cosmological constant, the scalar field is a dynamical variable, and density of the DE depends on time. The models can be distinguished by the choice of Lagrangian for the scalar field. Let us start from probably the simplest model of the DE of such type called quintessence.

The quintessence means the scalar field $\varphi$ in the potential $V(\varphi)$ minimally coupled to gravity, i.e. affected solely by the curvature of space-time, and limited by canonical form of kinetic energy. Action for such a field reads
\begin{equation}\label{Q_1}
S=\int d^4x\sqrt{-g}L=\int d^4x\sqrt{-g}\left[\frac12 g^{\mu\nu}\frac{\partial\varphi}{\partial x_{\mu}} \frac{\partial\varphi}{\partial x_{\nu}}-V(\varphi)\right],
\end{equation}
where $g={\rm det\ }g_{\mu\nu}$. Equation of motion for the scalar field is obtained by variation of the action w.r.t. the field
\begin{equation}\label{Q_2}
\frac1{\sqrt{-g}}\partial_{\mu}\left(\sqrt{-g}\ g^{\mu\nu}\frac{\partial\varphi}{\partial x_{\nu}}\right) =-\frac{d V(\varphi)}{d\varphi}.
\end{equation}
In flat Friedmannian Universe, i.e. for the FLRW metric, one obtains for the homogeneous field $\varphi(t)$
\begin{equation}\label{Q_3}
\ddot\varphi+3H\dot\varphi+V'(\varphi)=0.
\end{equation}
For the case of homogeneous field $\varphi(t)$ in locally Lorentzain frame, where the metric $g_{\mu\nu}$ can be replaced by the Minkowski one, one obtains the following expression for density and pressure of the scalar field
\begin{align}\label{Q_4}
\rho_\varphi=T_{00}=\frac12\dot\varphi^2+&V(\varphi);\\\nonumber
p_\varphi=T_{ii}=\frac12\dot\varphi^2-&V(\varphi).
\end{align}
The Friedmann equations for flat Universe filled with the scalar field, and the Klein-Gordon equation take on the form
\begin{align}\nonumber
H^2&=\frac13\left[\frac12\dot\varphi^2+V(\varphi)\right];\\\label{Q_5}
\dot H&=-\frac{\dot\varphi^2}2;\\\nonumber
\ddot\varphi&+3H\dot\varphi+\frac{dV}{d\varphi}=0.
\end{align}
Using (\ref{Q_4}) one obtains the EoS for the scalar field
\begin{equation}\label{Q_6}
w_\varphi=\frac{p_\varphi}{\rho_\varphi}=\frac{\dot\varphi^2-2V}{\dot\varphi^2+2V}.
\end{equation}
The EoS parameter $w_\varphi$ for the scalar field varies in the interval
\begin{equation}\label{Q_7}
-1\le w_\varphi\le1.
\end{equation}
The EoS for the scalar field can be conveniently presented as
\begin{equation}\label{Q_8}
w_\varphi(x)=\frac{x-1}{x+1};\quad x\equiv\frac{\dot\varphi^2/2}{V(\varphi)}.
\end{equation}
The function $w(x)$ monotonically grows from the minimal value $w_{min}=-1$ at $x=0$ to maximal asymptotic value $w_{max}$ at $x\to\infty$, which corresponds to $V=0$. In the slow-roll limit $x\ll1$ ($\dot\varphi^2\ll V(\varphi)$) the scalar field behaves as the cosmological constant $w_{\varphi}=-1$. It is easy to see that $\rho_\varphi=const$ in this case. In the other limit $x\gg1$ ($\dot\varphi^2\gg V(\varphi)$) (rigid matter) $w_\varphi=1$. In this case the energy density of the scalar field evolves as $\rho_\varphi\propto a^{-6}$. The intermediate situation $x\sim1$, $p\sim0$ corresponds to the case of non-relativistic matter.

The DP for the Universe filled with the scalar field is
\begin{equation}\label{Q_9}
q_\varphi=\frac{d}{dt}\left(\frac1H\right)-1=\frac{\dot\varphi^2-V}{\frac12\dot\varphi^2+V}.
\end{equation}
Let us consider now a new formalism for the analysis of scalar fields in flat isotropic and homogeneous cosmological models, proposed in \cite{harko}. Using the formalism, the basic evolution equation of the models can be reduced to a first order non-linear differential equation. Approximate solutions of this equation can be constructed in the limiting cases of the scalar field kinetic energy and potential energy dominance, respectively, as well as in the intermediate regime. In the following we will restrict our study to expansionary cosmological models, which satisfy the condition that the scale factor is a monotonically increasing function of time. For expanding cosmological models the condition $H>0$ is always satisfied.

From Eqs. (\ref{Q_5}) we obtain the Riccati type equation satisfied by $H$, of the form
\begin{equation}
\dot{H}=V-3H^{2}.  \label{Q_10}
\end{equation}
By substituting the Hubble function into the Klein-Gordon equation, we obtain the basic equation describing the scalar field evolution as
\begin{equation}
\ddot{\varphi}+\sqrt{3}\sqrt{\frac{\dot{\varphi}^{2}}{2}+V\left( \varphi \right) }\;
\dot{\varphi}+\frac{dV}{d\varphi }=0.  \label{Q_11}
\end{equation}
In order to deduce a basic equation describing the dynamics of the
scalar fields in the flat FLRW Universe, we define a new function $G(\varphi)$ by \[\sinh G(\varphi)\equiv\frac{\dot\varphi}{\sqrt{2V(\varphi)}},\] so that Eq. (\ref{Q_11})
becomes
\begin{equation}
\frac{dG}{d\varphi }+\frac{1}{2V}\frac{dV}{d\varphi }\coth G+\sqrt{\frac{3}{2}}=0.
\label{Q_14}
\end{equation}
As a function of time $G$ satisfies the equation
\begin{equation}
\frac{dG}{dt}=-\sqrt{2V\left( \varphi \right) }\sinh G\left[ \sqrt{\frac{3}{2}}+%
\frac{1}{2V\left( \varphi \right) }\frac{dV}{d\varphi }\coth G\right] .
\label{Q_15}
\end{equation}
Let us now turn directly to the question of our interest --- DP as a function of the scalar field. Using (\ref{Q_14}) one obtains
\begin{equation}\label{Q_18}
q(\varphi )=3\tanh ^{2}G(\varphi )-1=2-3\left( 1+\frac{\dot{\varphi}^{2}}{2V(\varphi )}\right) ^{-1}.
\end{equation}
The latter result evidently coincides with the Eq. (\ref{Q_9}).

The considered approach enables us to obtain a number of analytic expressions for DP in a wide class of scalar field potentials. Let us briefly cite below several examples (see \cite{harko} for details).

\paragraph{The exponential potential scalar field.}
If $V^{\prime }/V$ is a constant, that is, $V^{\prime }/V=\sqrt{6}\alpha
_{0}=\ const$, the scalar field self-interaction potential is of the
exponential form,
\begin{equation}
V=V_{0}\exp \left( \sqrt{6}\alpha _{0}\varphi \right) ,  \label{pp}
\end{equation}%
where $V_{0}$ is an arbitrary constant of integration, then Eq. (\ref{Q_14}) takes the form
\begin{equation}
\frac{dG}{d\varphi }+\sqrt{\frac{3}{2}}\left( \alpha _{0}\coth G+1\right) =0.
\label{dd}
\end{equation}%

For the case $\protect\alpha _0\neq \pm 1$ Eq.~(\ref{dd}) gives immediately
\begin{equation}
\sqrt{\frac{3}{2}}\left[ \varphi (G)-\varphi _{0}\right] =\frac{G-\alpha _{0}\ln
\left| \sinh G+\alpha _{0}\cosh G\right| }{\alpha _{0}^{2}-1}\,,
\label{expphi}
\end{equation}%
where $\varphi _{0}$ is an arbitrary constant of integration. The time
dependence of the physical parameters can be obtained from Eq.~(\ref{Q_15})
as
\begin{equation}
t(G)-t_{0}=-\frac{1}{\sqrt{3V_{0}}}\int {\frac{dG}{e^{\sqrt{3/2}\alpha
_{0}\varphi }\left( \sinh G+\alpha _{0}\cosh G\right) }}.  \label{exp1b}
\end{equation}

\begin{figure*}
\includegraphics[width=\textwidth]{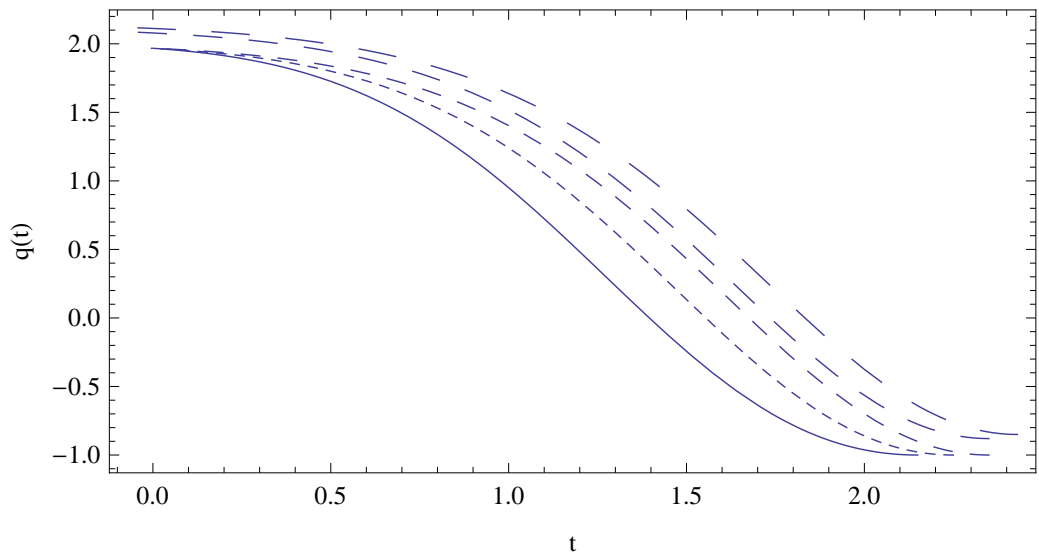}
\caption{Plot of the time variation of the DP of the Universe filled with an exponential potential scalar field  for different values of $\alpha _0$: $\alpha _0=1.5 $ (solid curve), $\alpha _0=2.5$ (dotted curve), $\alpha _0=3.5$ (short dashed curve), $\alpha _0=4.5$ (dashed curve), and $\alpha _0=5.5$ (long dashed curve), respectively. The arbitrary integration constants $\varphi _0$ and $V_0$ have been normalized so that $\exp \left(-\sqrt{3/2}\alpha _0\varphi _0\right)=\sqrt{3V_0}$.}
\label{exp4}
\end{figure*}

\begin{figure*}
\includegraphics[width=\textwidth]{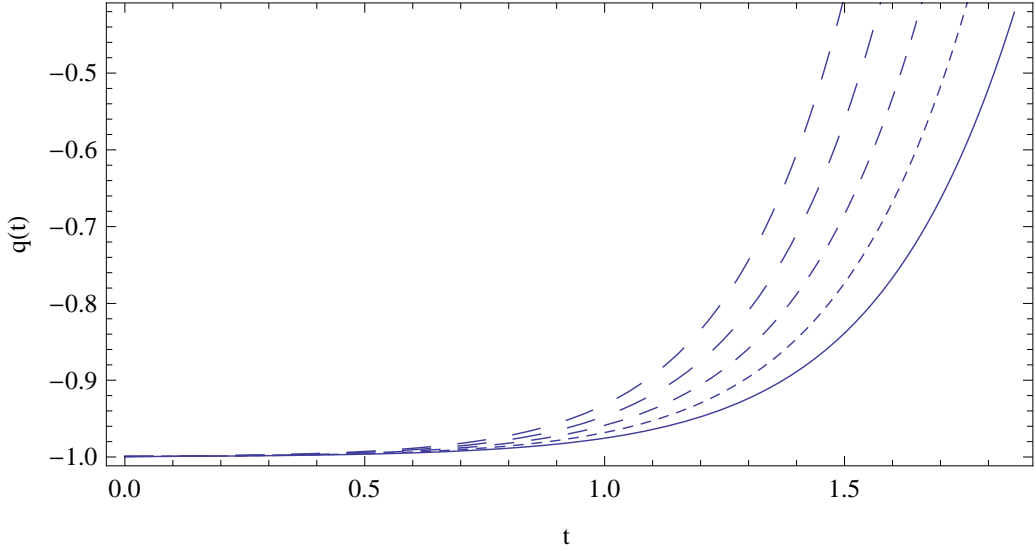}
\caption{Time variation of the DP of the Universe filled with a scalar field with a generalized hyperbolic cosine self-interaction potential for different values of $\alpha _1$: $\alpha _1=0.1 $ (solid curve), $\alpha _1=0.15$ (dotted curve), $\alpha _1=0.20$ (short dashed curve), $\alpha _1=0.25$ (dashed curve), and $\alpha _1=0.30$ (long dashed curve), respectively.}
\label{hyp4}
\end{figure*}

With the use of Eq.~(\ref{expphi}), we obtain the following integral
representation for the time $t$,
\begin{equation}  \label{exp2}
t(G)-t_{0}=-\frac{e^{-\sqrt{\frac32}\alpha _{0}\varphi _{0}}}{\sqrt{3V_{0}}}
\int {e^{\frac{\alpha _{0}}{1-\alpha _{0}^{2}}G}\left( \sinh
G+\alpha _{0}\cosh G\right) ^{\frac1{\alpha _{0}^{2}-1}}dG},
\end{equation}
where $t_{0}$ is an arbitrary constant of integration. Thus, Eqs.(\ref{Q_18}) and (\ref{exp2}) give a parametric representation of the time evolution of the DP, with $G$ taken as a parameter. It is represented, for different values of $\alpha _0$,
in Fig.~\ref{exp4}. For the considered range of the parameter $\alpha _0$, the Universe
starts its evolution from a decelerating phase, with $q>0$, but after a
finite interval it enters in an accelerated era, with $q<0$. In the limit of large times
the Universe enters a de Sitter type accelerated phase, with $q=-1$.
\paragraph{Generalized hyperbolic cosine type scalar field potentials.}
As a second example of an exact integrability of the evolution equation,
given by Eq.~(\ref{Q_14}), we consider the case in which the scalar field
potential can be represented as a function of $G$ in the form
\begin{equation}
\frac{1}{2V}\frac{dV}{d\varphi }=\sqrt{\frac{3}{2}}\;\alpha _{1}\,\tanh G,
\end{equation}%
where $\alpha _{1}$ is an arbitrary constant. With this choice, the
evolution equation takes the simple form
\begin{equation}
\frac{dG}{d\varphi }=\sqrt{\frac{3}{2}}\left( 1+\alpha _{1}\right) ,
\end{equation}%
with the general solution given by
\begin{equation}
G\left( \varphi \right) =\sqrt{\frac{3}{2}}\left( 1+\alpha _{1}\right) \left(
\varphi -\varphi _{0}\right) ,
\end{equation}%
where $\varphi _{0}$ is an arbitrary constant of integration.
The time dependence of the DP can be obtained in a parametric form (taking $\varphi$ as the parameter)
as
\begin{equation}
t-t_{0}=\frac{1}{\sqrt{2V_{0}}} \int \frac{d\varphi }{\cosh ^{\frac{\alpha _1}{%
1+\alpha _1}}\left[ \sqrt{\frac{3}{2}}\left( 1+\alpha _1\right) \left( \varphi
-\varphi _{0}\right) \right] \sinh \left[ \sqrt{\frac{3}{2}}\left( 1+\alpha _1
\right) \left( \varphi -\varphi _{0}\right) \right] }.
\end{equation}
\begin{equation}
q=3\tanh ^{2}\left[ \sqrt{\frac{3}{2}}\left( 1+\alpha _1\right) \left( \varphi
-\varphi _{0}\right) \right] -1.
\end{equation}
The time variation of the DP is represented for different values of $\alpha _1$, in Fig.~\ref{hyp4}.
In all of the considered models the Universe shows an expansionary,
accelerated behavior, starting with an initial value $q=-1$ of the
DP. The scalar field potential is increasing in time,
leading, in the long time limit, to accelerated expansions, with $q > -1$.

\paragraph{Power-law type scalar field potential.}
A simple solution of the gravitational field equations for a power-law type
scalar field potential can be obtained by assuming for the function $G$ the
following form
\begin{equation}
G=\mathrm{arccoth}\left( \sqrt{\frac{3}{2}}\frac{\varphi }{\alpha _{2}}\right)
,\qquad \alpha _{2}=\mathrm{constant}.
\end{equation}
With this choice of $G$, the Eq.~(\ref{Q_14}) immediately provides the
scalar field potential given by
\begin{equation}
V\left( \varphi \right) =V_{0}\left( \frac{\varphi }{\alpha _{2}}\right)
^{-2\left( \alpha _{2}+1\right) }\left[ \frac{3}{2}\left( \frac{\varphi }{%
\alpha _{2}}\right) ^{2}-1\right] ,  \label{mmm}
\end{equation}%
where $V_{0}$ is an arbitrary constant of integration. The time dependence
of the scalar field is given by a simple power law,
\begin{equation}
\frac{\varphi (t)}{\alpha _{2}}=\left[ \frac{\sqrt{2V_{0}}\left( \alpha
_{2}+2\right) }{\alpha _{2}}\right] ^{\frac{1}{\alpha _{2}+2}}\left(
t-t_{0}\right) ^{\frac{1}{\alpha _{2}+2}}.
\end{equation}%
The DP is given by
\begin{equation}
q=2\left( \frac{\varphi }{\alpha _{2}}\right) ^{-2}-1=
2\left[ \frac{\sqrt{2V_{0}}\left( \alpha _{2}+2\right) }{\alpha _{2}}\right] ^{-\frac{2}{\alpha _{2}+2}}
\left( t-t_{0}\right) ^{-\frac{2}{\alpha _{2}+2}}-1.
\end{equation}
\section{Chaplygin gas}
Multiple attempts to explain the observed accelerated expansion of the Universe in frames of GR stimulated consideration of a new type of matter as candidate for DE. The Chaplygin Gas (CG) realizes one of such possibilities. The EoS for CG is
\begin{equation}\label{cg_1}
p=-\frac A\rho.
\end{equation}
Desire to improve the description of the observed dynamics of the Universe led to multiple generalizations of the EoS (\ref{cg_1}).

The EoS for generalized Chaplygin gas (GCG) \cite{cg_1} is given by
\begin{equation}\label{cg_2}
p=-\frac A{\rho^\alpha},
\end{equation}
where $0\le\alpha\le1$. The additional parameter $\alpha$ widens possibilities of the model. Further modification of the EoS (\ref{cg_1}) includes a positive term linearly proportional to the density
\begin{equation}\label{cg_3}
p=B\rho-\frac A{\rho^\alpha},
\end{equation}
Here $A,B,\alpha$ are positive constants and $0\le\alpha\le1$. Such a model is called the modified Chaplygin gas (MCG). The cosmological models based on the EoS (\ref{cg_1}) and its numerous generalization of the type (\ref{cg_2}) and (\ref{cg_3}) share an attractive feature to describe substances with essentially different physical properties in frames of single EoS. So for instance the MCG model (\ref{cg_3}) for low densities describes a substance generating negative pressure while at high density and $B=1/3$ this equation describes a radiation dominated era.

Using the EoS (\ref{cg_3}), we can integrate the conservation equation to obtain
\begin{equation}\label{cg_4}
\rho_{MCG}(a)=\rho_0\left[A_S+\frac{1-A_S}{a^{3(1+B)(1+\alpha)}}\right]^{\frac1{1+\alpha}},
\end{equation}
where \[A_S\equiv\frac A{1+B}\frac1{\rho_0^{\alpha+1}},\]
$B\ne0$ and $\rho_0$ is an integration constant. In the considered model of Universe one have to introduce only the baryon component $\Omega_{b}$ (there is no necessity for introduction of the DE), as the Chaplygin gas represents an unified model of both the dark components. Using
\begin{equation}\label{cg_5}
q=\frac{\Omega_{tot}}2+\frac32\sum\limits_iw_i\Omega_i,
\end{equation}
for the case of spatially flat Universe one finds \cite{paul_thakur_beesham}
\begin{align}
q(z)&=\frac{a^{-3}\Omega_{b0}+\Omega_{MCG}(a)[1+3w_{MCG}(a)]}{2\left[{a^{-3}}\Omega_{b0}+\Omega_{MCG}(a)\right]}\\
\Omega_{MCG}(a)&=\Omega_{MCG0}[A_S+(1-A_S)a^{-3(1+B)(1+\alpha)}]^{\frac1{1+\alpha}}\label{cg_6}
\end{align}
The EoS parameter $w$ for the MCG can be found using its definition (\ref{cg_3})
\begin{equation}\label{cg_7}
w_{MCG}=\frac p\rho=B-\rho^{-(\alpha+1)}.
\end{equation}
Substitution of the equation (\ref{cg_4}) into (\ref{cg_7}) gives
\begin{equation}\label{cg_8}
w_{MCG}(a)=B-\frac{A_S(1+B)}{A_S+(1-A_S)a^{-3(1+B)(1+\alpha)}}.
\end{equation}
Let us now express parameters of the CG with the EoS (\ref{cg_1}) in terms of the cosmographic parameters. As we have seen above for small values of $a(t)$ density of the Chaplygin gas reduces to the dust density $\rho\propto a^{-3}$, and for large $a$ one gets the de Sitter Universe: $\rho=const$, $p=-\rho$. In between these two regimes one can use the approximation
\begin{equation}\label{scalar_5}\rho=\sqrt A+\frac{B}{\sqrt{2A}}a^{-6}.\end{equation}
Thus $\sqrt A$ plays the role of a cosmological constant. We insert this to the Friedmann equation \begin{equation}\dot a^2+k=\frac13\rho a^2+\frac13\Lambda a^2\end{equation}
with $\Lambda$ and follow the procedure of eliminating the constants by differentiation. This leads to an approximate constraint
\begin{equation}\label{scalar_6}s+5(q+j)+qj=0.\end{equation}
Analogous procedure for the generalized Chaplygin gas (\ref{cg_2}) leads to the equation \cite{dunajski_gibbons}
\begin{equation}\label{scalar_7}s+(3\alpha+2)(q+j)+qj=0.\end{equation}
in agreement with
\begin{equation}s+2(q+j)+qj=0.\end{equation}
and (\ref{scalar_6}) when $\alpha=0$ and $\alpha=1$ respectively. If we want to exclude the parameter $\alpha$ from the latter equation we must take one more derivative of the Friedmann equation and introduce an additional cosmological parameter \[l=\frac1a\frac{d^5a}{dt^5}\left(\frac1a\frac{da}{dt}\right)^{-5}.\] As a result one obtains
\begin{equation}\label{scalar_8}
-2qs-2jq^2-lq-2sj-3sq^2-j^2q-lj+s^2-3q^2j-qsj+j^3-2j^2q^2=0.
\end{equation}
This constraint is again approximate and is valid only in the regime where the higher order terms in (\ref{scalar_5}) can be dropped.
\section{$K$-essence}
As we have seen above, in the quintessence model the required dynamical behavior is attained due to appropriate choice of the scalar field potential. Another DE model called the $k$-essence \cite{k_1,k_2,k_3,k_4} is realized by the scalar field with modified kinetic term.

Let us introduce the quantity \[X\equiv\frac12g^{\mu\nu}\frac{\partial\varphi}{\partial x_{\mu}}\frac{\partial\varphi}{\partial x_{\nu}}\] and consider the scalar field action of the form
\begin{equation}\label{k_e_1}
S=\int d^4x\sqrt{-g}F(\varphi,X),
\end{equation}
where $F$ is generally speaking arbitrary function of the variables $\varphi$ and $X$. Let us first consider the $\varphi$-independent case \cite{k_5}
\begin{equation}\label{k_e_2}
S=\int d^4x\sqrt{-g}F(X).
\end{equation}
Using the energy momentum tensor for a perfect fluid we obtain that the $k$-essence energy density $\rho$ and pressure $p$ are
\begin{equation}\label{k_e_3}
\rho=2XF_X-F,\quad p=F.
\end{equation}
We will assume that the energy density is positive so that $2XF_X-F>0$. The EoS is
\begin{equation}\label{k_e_4}
w=\frac p\rho=\frac F{2XF_X-F}.
\end{equation}
One can use  equation (\ref{k_e_4}) to express the condition $w>-1$ ($w<-1$) as a condition on  the function $F(X)$. We need to consider the two possibilities $F>0$ and $F<0$ separately. In the first case, demanding the energy density be positive immediately implies $w>0$. For $F < 0$, a positive energy density means that also $2XF_X/F<1$ so
\begin{equation}\label{k_e_5}
w=\frac{-1}{1-2XF_X/F}>-1,
\end{equation}
when $F_X>0$. Therefore \cite{k_6}
\begin{align}
\nonumber F&<0, & F_X>0&\Rightarrow w>-1;\\
\label{k_e_6} F&<0, & F_X<0&\Rightarrow w<-1;\\
\nonumber F&>0,& &\Rightarrow w>0.
\end{align}
The latter condition excludes the possibility of the accelerated expansion.

Next, we consider the case $F(\varphi,X)$. The non-canonical scalar field Lagrangian density is given by Fang et al. \cite{Fang},
\begin{equation}\label{ldensity}
F(\varphi,X) = \Gamma(X) - V(\varphi),
\end{equation}
where $V(\varphi)$ is a self-interacting potential for the scalar field $\varphi$, $\Gamma(X)$ is an arbitrary function of $X$, which is defined as $X={\dot{\varphi}}^2/2.$ Following the article \cite{1503.06280} let us consider a Lagrangian density of the following form
\begin{equation}\label{L}
F(\varphi,X) = X^2 - V(\varphi),\hspace{5mm}X = \frac{1}{2}{\dot{\varphi}}^{2}.
\end{equation}
In this case, the energy density and pressure associated with this Lagrangian density read
\begin{equation}\label{rhophi}
\rho_{\varphi} = \frac{3}{4}{\dot{\varphi}}^4 + V(\varphi);
\end{equation}
\begin{equation}
p_{\varphi} = \frac{1}{4}{\dot{\varphi}}^4 - V(\varphi),
\end{equation}
and Friedman equations with conservation equations take on the form
\begin{equation}\label{eq1}
3H^{2} = {\rho}_{m} + \frac{3}{4}{\dot{\varphi}}^4 + V(\varphi);
\end{equation}
\begin{equation}\label{eq2}
2{\dot{H}} + 3H^{2} = -\frac{1}{4}{\dot{\varphi}}^4 + V(\varphi);
\end{equation}
\begin{equation}\label{eq3}
{\dot{\rho}}_{\varphi} + 3H(\rho_{\varphi} + p_{\varphi}) = 0;
\end{equation}
\begin{equation}\label{eq4}
{\dot{\rho}}_{m} + 3H{\rho}_{m} = 0.
\end{equation}
Among this four equations (equations (\ref{eq1})-(\ref{eq4})), only three are independent equations with four unknown parameters $H$, $\rho_{m}$, $\varphi$ and $V(\varphi)$ and for this reason we still have freedom to choose one parameter to close the above system of equations. As for this parameter we can choose parameter of DE EoS. Parametrization of EoS parameter as,
\begin{equation}
w_{\varphi}(z) = \frac{p_{\varphi}}{\rho_{\varphi}}=w_{0} + w_{1}f(z),
\end{equation}
where $w_{0}$, $w_{1}$ are real numbers and $f(z)$ is a function of redshift $z$.

The most popular parametrization Chevallier-Polarski-Linder (CPL) parametrization  is one of the most popular one and is given by
\begin{equation}\label{cpl}
w_{\varphi}(z) = w_{0} + w_{1}(1 - a) = w_{0} + w_{1}\left(\frac{z}{1+z}\right),
\end{equation}
where $z = 1/a - 1$ is the redshift,  $w_{0}$  represents the current value of $w_{\varphi}(z)$ and the second term represents variation of the EoS parameter w.r.t. redshift. Note that this parametrization has the advantage of giving finite $w_{\varphi}$ in the entire range, $0 < z < \infty$.

\begin{figure}[!h]
\includegraphics[width=0.45\textwidth]{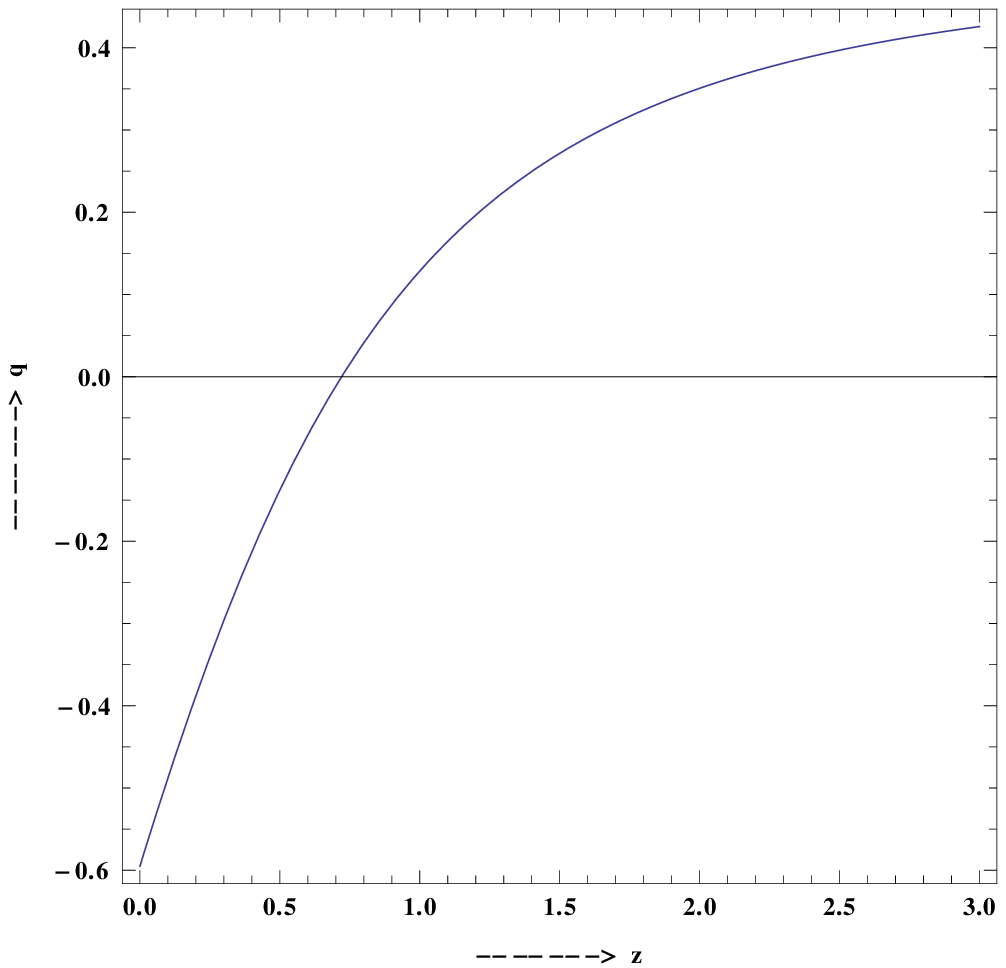}
\includegraphics[width=0.45\textwidth]{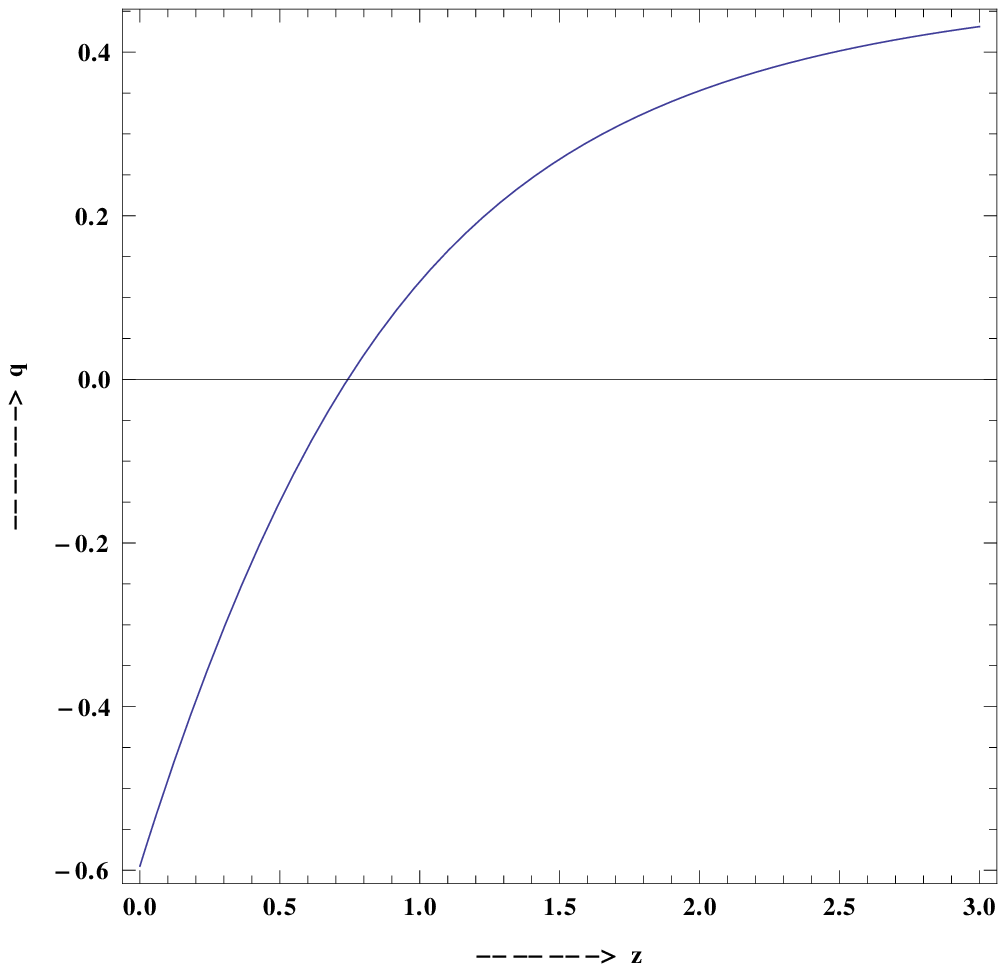}\\
\includegraphics[width=0.45\textwidth]{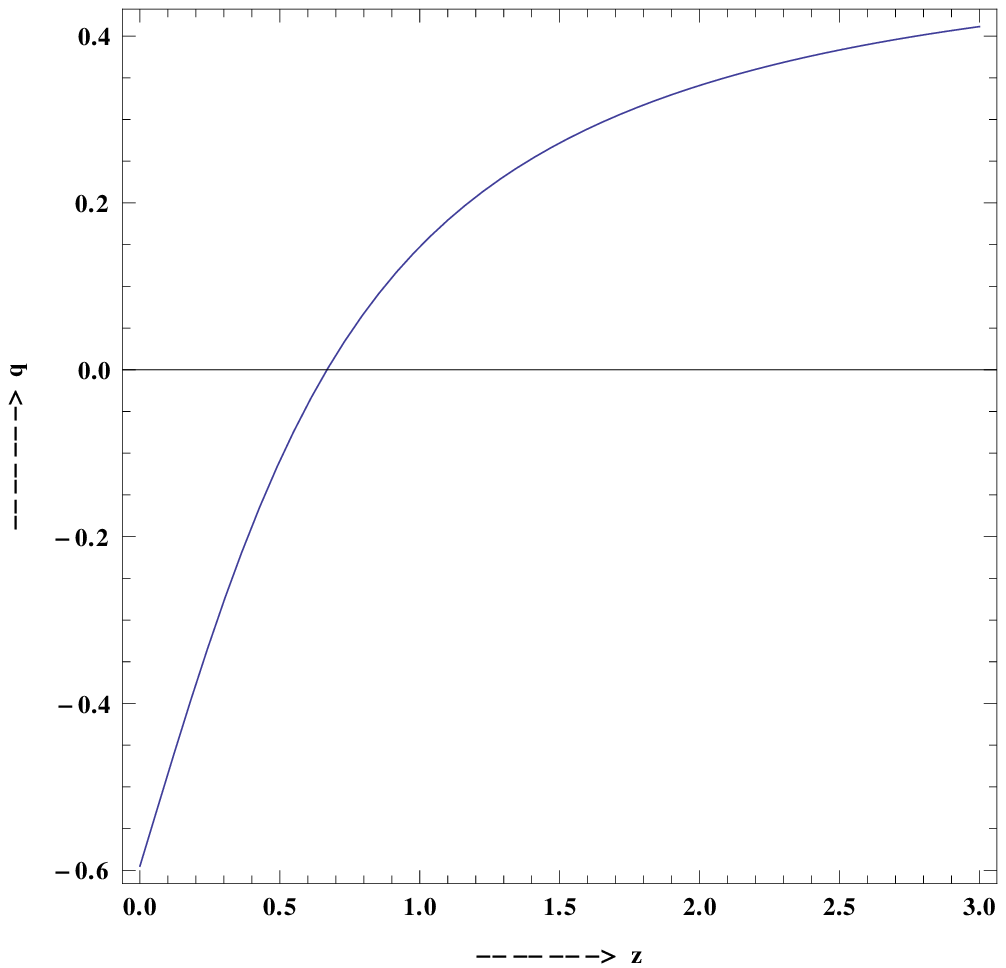}
\includegraphics[width=0.45\textwidth]{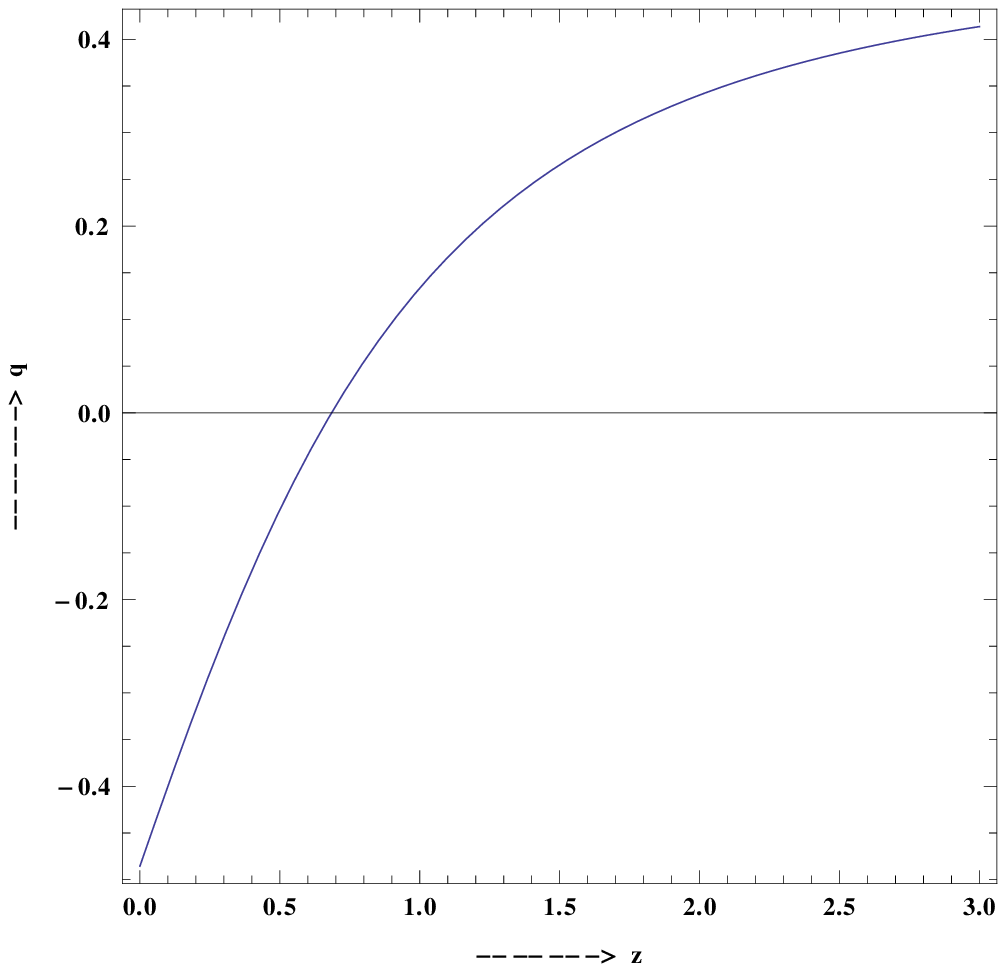}
\caption{\normalsize{\em a) Top left: $q(z)$ for CPL parametrization.  b) Top right: $q(z)$ for extended CPL parametrization. c) Bottom left: $q(z)$ for Barboza-Alcaniz parametrization. d) Bottom right: $q(z)$ for Generalized Chaplygin Gas parametrization. For all figures $\kappa = \Omega_{m0}/\Omega_{\varphi 0}= 0.27/0.73$, $w_{0} = -1$ and $w_{1} = 0.2$}}
\label{figq}
\end{figure}

The DP for this model takes the following form
\begin{equation}
q(z) = \frac{1}{2} + \frac{3}{2}\left[\frac{w_{0} + w_{1}\left(\frac{z}{1+z}\right)}{1 + \kappa (1 + z)^{(3 - 3\alpha_{1})}  e^{\left(\frac{3w_{1}z}{1+z}\right)}}\right]
\end{equation}
where \[\kappa = \frac{\rho_{m0}}{\rho_{\varphi 0}} = \frac{\Omega_{m0}}{\Omega_{\varphi 0}}.\] If $q<0$ the model accelerates while $q>0$ indicates deceleration of the universe. From the figure (\ref{figq}a), we have seen that $q$ decreases from positive to negative value by suitable choices of model parameters \cite{1503.06280}.

However, the model presented here is restricted because $\omega_{\varphi}(z)$ diverges when $z\rightarrow -1$, i.e. this model cannot predict about the future evolution. So, the model can nicely describe the evolution history of the universe in the past and near future up to $z\geq -1$ but can not predict about the evolution beyond that limit.

Recently, Jassal et al. {\cite{jbp}} extended the above parametrization to a more general case:
\begin{equation}\label{eqnjbppara}
\omega_{\varphi}(z) = \omega_{0} + \omega_{1}\frac{z}{(1+z)^{p}}.
\end{equation}
In this model the DP is expressed as
\begin{equation}
q(z) = \frac{1}{2} + \frac{3}{2}\left[\frac{w_{0} + w_{1}\frac{z}{(1+z)^2}}{1 + \kappa (1 + z)^{- 3w_{0}}  e^{-\left(\frac{3w_{1} z^2}{2(1+z)^2}\right)}}\right]
\end{equation}

Figure (\ref{figq}b) shows the plot of $q(z)$ as a function of $z$. This plot clearly shows the transition of $q$ from the decelerating to the accelerating regime at $z \sim 0.8$. In graph, the resulting cosmological scenarios are in good agreement with observations. For the JBP model also, $\omega_{\varphi}(z)$ diverges as $z\rightarrow-1$ and thus future evolution can not be predicted.

The next parametrization was proposed by Barboza et al. \cite{ba}, which has the following functional form
\begin{equation}
w_{\varphi}(z) = w_{0} + w_{1}\frac{z(1+z)}{1+z^2},
\end{equation}
where $w_{\varphi}(z=0)=w_{0}$ (the present value of the EoS parameter), \[w_{1} = \frac{dw_{\varphi}}{dz}|_{z=0}\] (which measures the variation of the EoS parameter with $z$), $w_{\varphi}(z=\infty) = w_{0} + w_{1}$ and the EoS parameter reduces to $w_{\varphi}(z) = w_{0} + w_{1}z$ at the low redshift ($z\ll1$). It is remarkable that the BA parametrization does not diverge like CPL model when $z\rightarrow -1$ and it remains bounded within the entire range $z\in [-1, \infty )$.\\
In this model the DP $q(z)$ can be expressed as
\begin{equation}
q(z) = \frac{1}{2} + \frac{3}{2}\left[\frac{w_{0} + w_{1}\frac{z(1+z)}{1+z^2}}{1 + \kappa (1 + z)^{- 3w_{0}}(1+z^2)^{-{\frac{3w_{1}}{2}}}}\right].
\end{equation}

Figure (\ref{figq}c) shows the evolution of the DP with redshift $z$. We have observed from figure (\ref{figq}c) that the evolution of the universe is in accelerating phase ($q<0$) at present epoch.

As we have seen in the previous section, the generalized Chaplygin gas behaves like dark matter (DM) in the past and it behaves like cosmological constant at present. Assuming that the Universe contains both the DM and the GCG, we also take into account the possibility that the $k$-essence $\varphi$ plays the role of GCG to explore the late time cosmic scenarios.

By inserting equation (\ref{cg_2}) into the energy conservation equation (\ref{eq3}), one finds that the density  of the scalar field $\varphi$ evolves as
\begin{equation}\label{rhogcg}
\rho_{\varphi}(z) = {\left[A + B(1+z)^{3(1+\alpha)}\right]}^{\frac{1}{(1+\alpha)}},
\end{equation}
where $B$ is an integration constant. Equation (\ref{rhogcg}) can be re-written in the following form
\begin{equation}
\rho_{\varphi}(z) = \rho_{\varphi 0}{\left[A_{s} + (1-A_{s})(1+z)^{3(1+\alpha)}\right]}^{\frac{1}{(1+\alpha)}},
\end{equation}
where, for simplicity, we have defined \[A_{s}=\frac{A}{A+B}\] and $\rho_{\varphi 0} = {\left(A + B\right)}^\frac{1}{(1+\alpha)}$ is the present value of the energy density of the GCG. To ensure the finite and positive value of $\rho_{\varphi}$ we need $-1< \alpha \leq 1$ and $0\le A_{s}\leq 1$.\\
\par In this case, the Hubble parameter is given by
\begin{equation}
H^2 = H^2_{0}{\left[ \Omega_{m0}(1+z)^{3} + \Omega_{\varphi 0}{\left(A_{s} + (1-A_{s})(1+z)^{3(1+\alpha)}\right)}^{\frac{1}{(1+\alpha)}}\right]}.
\end{equation}
The corresponding expression for the EoS parameter is given by
\begin{equation}
\omega_{\varphi}(z)=-\frac{A_{s}}{A_{s} + (1- A_{s})(1+z)^{3(1+\alpha)}}.
\end{equation}
Like earlier mentioned three models, the EoS parameter of the GCG also depends on two independent model parameters ($A_{s}$ and $\alpha$) along with redshift $z$. At present epoch, the above EoS parameter becomes, $\omega_{\varphi}(z=0)= -A_{s}$. In this case, it is very interesting to notice that the GCG will behave like pure cosmological constant when we put $A_{s}=1$.\\
\par The DP $q$ can be written as
\begin{equation}
q(z) = \frac{1}{2} + \frac{3}{2}{\left[-\frac{\frac{A_{s}}{A_{s} + (1- A_{s})(1+z)^{3(1+\alpha)}}}{1+\frac{\kappa(1+z)^3}{{\left[A_{s} + (1-A_{s})(1+z)^{3(1+\alpha)}\right]}^{\frac{1}{(1+\alpha)}}}}\right]},
\end{equation}
where \[\kappa = \frac{\rho_{m0}}{\rho_{\varphi 0}} = \frac{\Omega_{m0}}{\Omega_{\varphi 0}}.\]

Figure (\ref{figq}d) shows the evolution of $q(z)$ vs. redshift $z$ for $A_{s} = 0.9$ and $\alpha = -0.5$. In fact, at low redshift, the transition of $q(z)$ from decelerating to accelerating regime depends upon the choice of the parameters $A_{s}$ and $\alpha$.
\section{Polytropic model}
Following Chavanis \cite{polytrop_1,polytrop_2,polytrop_3} we construct models of Universe with a EoS having a linear component and a polytropic component,
\begin{equation}\label{poly_1}
p=\alpha\rho+k\rho^{1+1/n}.
\end{equation}
Concerning the linear EoS, we assume $-1\le\alpha\le1$. As for the polytropic component, we consider a wide range of parameters.

With positive index $n>0$, the polytropic component dominates the linear component when the density is high. These models describe the early Universe. Conversely, when $n<0$, the polytropic component dominates the linear component when the density is low. If $1+\alpha+k\rho^{1/n}>0$, the EoS parameter is $w>-1$, which corresponds to the quintessential phase. The opposite case is $1+\alpha+k\rho^{1/n}<0$, leading to a phantom Universe phase. Below we consider behavior of the DP for different values of the parameters $\alpha, n, k$.

We start consideration from the case $1+\alpha+k\rho^{1/n}>0$, $n>0$, when the equation
\begin{equation}\label{poly_2}
\dot\rho+3H\rho(1+\alpha+k\rho^{1/n})=0
\end{equation}
has the following solution
\begin{equation}\label{poly_3}
\rho=\frac{\rho_*}{\left[(a/a_*)^{3(1+\alpha)/n}\mp1\right]^n},
\end{equation}
where $\rho_*\equiv[(\alpha+1)/|k|]^n$ and $a_*$ is a constant of integration. The upper sign corresponds to $k>0$ (repulsive self-interaction) and the lower sign corresponds to $k<0$ (attractive self-interaction).

We can rewrite the EoS (\ref{poly_1}) as $p=w(t)\rho$ with
\begin{equation}\label{poly_4}
w(t)=\alpha\pm(\alpha+1)\left(\frac\rho{\rho_*}\right)^{1/n}.
\end{equation}
For a flat Universe
\begin{equation}\label{poly_5}
q(t)=\frac{1+3w(t)}2
\end{equation}
and consequently
\begin{equation}\label{poly_6}
q(t)=\frac{1+3\alpha}2\pm\frac32(\alpha+1)\left(\frac\rho{\rho_*}\right)^{1/n}.
\end{equation}
We define a critical density $\rho_c$ and  a critical scale factor $a_C$, corresponding to a possible inflexion point $q=0$ in the curve a(t)
\begin{equation}\label{poly_7}
\rho_c=\left[\mp\frac{1+3\alpha}{3(1+\alpha)}\right]^n, \quad a_c=\left[\mp\frac2{1+3\alpha}\right]^{n/[{3(1+\alpha)}]}.
\end{equation}
Let us now analyze behavior of the DP for positive and negative values of the parameter $k$. We first assume $k>0$.  In this case  the density is defined only for $a>a_*$.When $a\to a_*$, $q\to\infty$. When $a\to\infty$,  $q\to(1+3\alpha)/2$. For $\alpha>-1/3$, the Universe is always decelerating ($q>0$). For $\alpha<-1/3$, the Universe is decelerating when $a_*<a<a_c$ and accelerating when $a>a_c$.

We now assume $k<0$.  In this case the density is defined for all $a$. When $a\to0$, $q\to-1$; when $a\to\infty$,  $q\to(1+3\alpha)/2$. For $\alpha<-1/3$, the Universe is always accelerating ($q<0$). For $\alpha>-1/3$, the Universe is accelerating when $a<a_c$ and decelerating when $a>a_c$.

Let us briefly analyze the case $1+\alpha+k\rho^{1/n}>0$, $n<0$. Dependence of the density on the scale factor is still described by the formula (\ref{poly_2}), however now for $k>0$ the density is defined only for $a<a_*$.

In this case again the DP in a flat Universe and inflexion point are defined by Eqs. (\ref{poly_6}), (\ref{poly_7}). We first assume $k>0$. When $a\to0$, $q\to(1+3\alpha)/2$, when $a\to a_*$,  $q\to+\infty$. For $\alpha>-1/3$, the Universe is always decelerating. For $\alpha<-1/3$, the Universe is accelerating when $a<a_c$ and decelerating when $a_c<a<a_*$.

We now assume $k<0$. When $a\to0$, $q\to(1+3\alpha)/2$, when $a\to\infty$, $q\to-1$. For $\alpha>-1/3$, the Universe is always decelerating. For $\alpha<-1/3$, the Universe is accelerating when $a<a_c$ and decelerating when $a>a_c$.

As a conclusion of the section we analyze the DP in the phantom Universe $1+\alpha+k\rho^{1/n}\le0$, $w(t)\le-1$. In this case the Friedmann equation (\ref{poly_2}) can be integrated  into
\begin{equation}\label{poly_8}
\rho=\frac{\rho_*}{\left[1-(a/a_*)^{3(1+\alpha)/n}\right]^n},
\end{equation}
where, as before, $\rho_*\equiv[(\alpha+1)/|k|]^n$ and $a_*$ is a constant of integration. For $n>0$, the density  is defined only when $a<a_*$.

The DP is defined by Eqs. (\ref{poly_4}), (\ref{poly_5})
\begin{equation}\label{poly_9}
q(t)=\frac{1+3\alpha}2-\frac32(\alpha+1)\left(\frac\rho{\rho_*}\right)^{1/n}.
\end{equation}
A phantom Universe is always accelerating since $q\le-1<0$.  For $n>0$, $q\to-1$ when $a\to0$ and $q\to-\infty$ when $a\to a_*$. For $n<0$, $q\to-\infty$ when $a\to a_*$ and $q\to-1$ when $a\to+\infty$.
\section{Tachyon}
The tachyon of string theory may be described \cite{tachy_gibbons} by effective field theory describing some sort of tachyon condensate which in flat space has a Lagrangian density
\begin{equation}\label{tachy_1}
L=-V(T)\sqrt{1+\eta^{\mu\nu}\partial_\mu T\partial_\nu T }.
\end{equation}
where $T$ is the tachyon field, $V(T)$ is the tachyon potential and $\eta_{\mu\nu}=diag(-,+,+,+)$ is the metric of Minkowski spacetime.  The tachyon potential $V(T)$ has a positive maximum at the origin and has a minimum at $T=T_0$ where the potential vanishes. In Minkowski spacetime the rolling down towards its minimum value is described by a spatially homogeneous but time-dependent solution obtained from the Lagrangian density
\begin{equation}\label{tachy_2}
L=-V\sqrt{1-\dot T^2}.
\end{equation}
During the rolling the Hamiltonian density
\begin{equation}\label{tachy_3}
\rho_H=\frac{V(T)}{\sqrt{1-\dot T^2}}
\end{equation}
has a constant value $E$. Thus
\begin{equation}\label{tachy_4}
\dot T=\sqrt{1-\frac{V^2}{E^2}}.
\end{equation}
As $T$ increases $V(T)$ decreases and $\dot T$ increases to attain its maximum value of $1$ in infinite time as $T$ tends to infinity. Note that as explained in \cite{tachy_sen} the tachyon field behaves like a fluid of positive energy density
\begin{equation}\label{tachy_5}
\rho=\frac{V(T)}{\sqrt{1-\dot T^2}}
\end{equation}
and negative pressure
\begin{equation}\label{tachy_6}
p=-V(T)\sqrt{1-\dot T^2}.
\end{equation}
Thus
\begin{equation}\label{tachy_7}
w=\frac p\rho=-(1-\dot T^2)
\end{equation}
and therefore, $-1\le w\le0$. Note that both the Weak Energy Condition, $\rho>0$ and Dominant Energy Condition, $\rho\ge|p|$ hold. However because
\begin{equation}\label{tachy_8}
\rho+3p=-\frac{-2V(T)}{\sqrt{1-\dot T^2}}\left(1-\frac32\dot T^2\right),
\end{equation}
the Strong Energy Condition fails to hold for small $|\dot T|$  but does hold for large $|\dot T|$.

The discussion above has neglected the gravitational field generated by the tachyon condensate. To take it into account we use the Lagrangian density
\begin{equation}\label{tachy_9}
L=\sqrt{-g}\left(\frac{R}{16\pi G}-V(T)\sqrt{1+g^{\mu\nu}\partial_\mu T\partial_\nu T }\right).
\end{equation}
In this case the expressions (\ref{tachy_5}) and (\ref{tachy_6}) for the density and pressure remain valid and thus the Friedman equations are
\begin{equation}\label{tachy_10}
\frac{\dot a^2}{a^2}+\frac k{a^2}=\frac{8\pi G}3\frac{V(T)}{\sqrt{1-\dot T^2}};
\end{equation}
\begin{equation}\label{tachy_11}
\frac{\ddot a}{a}=\frac{8\pi G}3\frac{V(T)}{\sqrt{1-\dot T^2}}\left(1-\frac32\dot T^2\right).
\end{equation}
It follows from (\ref{tachy_4}) that as $T$ increases $V(T)$ increases too, but $\dot T$ could decrease since $E$ decreases. In any case $\dot T$ remains positive and so $T$ increases monotonically to its maximum value of $1$. It follows from (\ref{tachy_10}) that if $k\le0$, then $\dot a$ will always be positive. This is because the Weak Energy Condition holds, $\rho>0$. From the Raychaudhuri equation (\ref{tachy_11}) it follows that if $|\dot T|<2/3$ the scale factor initially accelerates ($\ddot a>0$), but then, when $\dot T$ exceeds $\sqrt{2/3}$, the acceleration will stop and deceleration will start. If the Universe is flat ($k=0$), then $a(t)\to\rm constant$. In the hyperbolic Universe ($k=-1$) the scale factor increases linearly with time ($a\to t$). In both cases the final state of the Universe is flat, the case $k = -1$ being the Milne model. In the spherical case ($k=1$) the Universe will re-collapse. The possibility of cosmic acceleration arises from the positive potential $V(T)$.
\section{Chameleon fields}
In all the above considered dynamical models of the DE (quintessence, $k$-essence, phantom fields) the scalar field had only the self-interaction, described by the potential $V(\varphi)$. Absence of interaction with other components in the Universe, firstly, looks unnatural, and secondly, it limits possibilities of the model. However inclusion of the interaction (the procedure is rather simple from theoretical point of view) immediately faces the following fundamental difficulty.

In a local part of Universe (say the Solar System) available for precision measurements the theory is agreed with the experiment (observations) by introduction of four fundamental forces: the strong, the weak, the electromagnetic and the gravitational ones. Introduction of a new type of interaction automatically leads to appearance of the ''fifth force'' which was not yet observed. The latter fact imposes strict constraints on the interaction of the scalar field with matter: either the interaction must be considerably weaker than the gravitational one, or quanta of the field must be too heavy, i.e. the interaction must be short-range.

A natural question arises: is it possible to build a model of DE in form of the scalar field interacting with matter which would not contradict the equivalence principle well verified on scales of the Solar system? Recall that the DE denotes any substance capable to provide the accelerated expansion of Universe. Positive answer to the above posed question is given by the so-called chameleon models of scalar field \cite{cham1,cham2,cham3,cham4,cham5,cham6}.

The chameleon scalar fields are scalar fields coupled to matter with intensity of order of the gravitational forces (or even higher) and mass depending on density of the environment. On the cosmological scales where density is negligible this fields are ultralight. On the contrary, for example in vicinity of the Earth where density is essentially higher the fields acquire sufficiently great mass. In other words, properties of the fields including its value vary along with changes in density of the environment. That is why they are called the chameleon fields.

Note that introduction of forces depending on density presents a common tendency in physics. Let us cite the following example. In the sixties in order to calculate nuclear characteristics by the Hartree-Fock method different types of effective nucleon-nucleon interactions were investigated. It turned out that no one among the considered potentials led to consistent quantitative description of the nucleus. The problem was solved only after introduction of the effective interaction between the nucleons, which depended on density.

Action for the chameleon field represents sum of Hilbert-Einstein action, scalar field action and a term describing the interaction of the scalar field with matter
\begin{equation}\label{cham_1}
S=\int\left[\frac R2+\frac12\varphi_{,\mu}\varphi^{,\mu}-V(\varphi)+f(\varphi)L_m\right]\sqrt{-g}d^4x.
\end{equation}
Unlike the usual Einstein-Hilbert action, the matter Lagrangian $L_m$ is modified as $f(\varphi)L_m$, where $f(\varphi)$ is an analytic function of $\varphi$. This term brings about the nonminimal interaction between the cold DM and chameleon field.

For a spatially flat FLRW metric variation of the action (\ref{cham_1}) w.r.t. the metric tensor components leads to the following Friedmann equations
\begin{align}
\label{cham_2}3H^2=\rho_mf&+\frac12\dot\varphi^2+V(\varphi);\\
\label{cham_3}H^2+2\frac{\ddot a}{a}=&-\frac12\dot\varphi^2+V(\varphi).
\end{align}
Variation of the action (\ref{cham_1}) w.r.t. field $\varphi$ results in the equation of motion for the scalar field
\begin{equation}\label{cham_4}
\ddot\varphi+3H\dot\varphi+V'=-\rho_mf',
\end{equation}
where a prime indicates differentiation w.r.t. $\varphi$.

From the equations (\ref{cham_2}-\ref{cham_4}) one can obtain the modified conservation equation
\begin{equation}\label{cham_5}
\frac d{dt} (\rho_m f)+3H\rho_m f=\rho_mf'\frac{df}{dt},
\end{equation}
and for the matter density one obtains
\begin{equation}\label{cham_6}
\rho_m=\frac{\rho_0}{a^3}.
\end{equation}
It means that in spite of appearance of interaction with the chameleon field the matter density evolution remains unaffected.

Using the expressions (\ref{cham_2}) and (\ref{cham_3}), we obtain expressions for $\rho_{eff}$ and $p_{eff}$
\begin{align}
\rho_{eff}&\equiv\rho_{m}f+\frac{1}{2}\dot{\varphi}^{2}+V(\varphi)\equiv \rho_{ch}+\rho_{de},\label{roef}\\
p_{eff}&\equiv\frac{1}{2}\dot{\varphi}^{2}-V(\varphi)\equiv p_{de}, \label{pef}
\end{align}
where $de$ and $ch$ refer to the standard DE and to the chameleon energy respectively, and $p_{eff}=w_{eff}\rho_{eff}$. Here we introduce the so-called ''chameleon matter density'' as $\rho_{ch}=\rho_m f$. It worth noting that the density of chameleon field, as can be seen from (\ref{cham_4}), is characterized by value of $\varphi$ and the potential $V(\varphi)$, while $\rho_{ch}$ in its turn describes the DM density interacting with the chameleon field. The conservation equations for the DE and the chameleon field which interact with matter separately take on the form
\begin{eqnarray}
&&\dot{\rho_{ch}}+3H\rho_{ch}=Q, \label{roef1}\\
&&\dot{\rho_{de}}+3H(1+\omega_{de})\rho_{de}=-Q,\label{pef1}
\end{eqnarray}
where $Q$ is the interaction source and $p_{de}=w_{de}\rho_{de}$. Comparing (\ref{cham_5}) and (\ref{roef1}) one finds that \[Q= \frac{1}{4}\rho_{m}\dot{\varphi}f'=\frac{1}{4}\rho_{ch}\frac{\dot f}{f}.\] Introducing function \[g(t,\varphi,\dot{\varphi})\equiv \frac{\dot{f}(\varphi)}{f(\varphi)}\] we rewrite $Q$ as
\begin{equation}\label{q4}
Q=\frac{1}{4}g\rho_{ch}=\frac{1}{8}g(\dot{\varphi}^{2}(-1+\omega_{eff})-2V(1+\omega_{eff})).
\end{equation}
From equation (\ref{q4}) one observes that $t=t_{cross}$ when $\omega_{eff}=-1 $, then
\begin{eqnarray}
Q=-\frac{1}{4}g\dot{\varphi}^{2}\label{q41}.
\end{eqnarray}
Current value of the function $g$ directly affects interaction between the components, for instance the interaction is absent at $g=0$.

The equation (\ref{q4}) can be rewritten in the form
\begin{equation}\label{q4_q}
Q=\frac{1}{4}g\rho_{ch}=\frac{1}{12}g(\dot{\varphi}^{2}(2+q)-2V(1+q)).
\end{equation}

If we consider ratio of the chameleon matter density to that of the DE $r\equiv{\rho_{ch}}/{\rho_{de}}$, we then obtain
\begin{eqnarray}\label{eta1}
1+\frac{1}{r}=\frac{3H^{2}}{\rho_{ch}}.
\end{eqnarray}
From equations (\ref{eta1}) and (\ref{roef1}) one finds
\begin{eqnarray}\label{eta2}
\frac{\dot{r}}{(1+r)^{2}}+\frac{3rH}{1+r}(1+\frac{2}{3}\frac{\dot{H}}{H^{2}})=\frac{Q}{3H^{2}}.
\end{eqnarray}
In addition, using \[q=-1-\frac{\dot{H}}{H^2},\] we obtain
\begin{eqnarray}\label{eta3}
\frac{\dot{r}}{(1+r)^{2}}- \frac{rH}{1+r}(1+2q)=\frac{Q}{3H^{2}}.
\end{eqnarray}
Using equation (\ref{q4}) now we can find expression for $q$:
\begin{equation}\label{qq}
    q=\frac{2(-18H^3r(1+r)+18H^2\dot{r}+g(1+r)^2(V-\dot{\varphi}^2))}{(1+r)(72H^3r-g(1+r)(2V-\dot{\varphi}^2))}.
\end{equation}

Having four unknown functions ($a(t),\ H(t),\ \varphi(t),\ \rho_m(t)$) and only three independent equations (\ref{cham_2},\ref{cham_3},\ref{cham_4}) we are forced to introduce an additional ansatz. In the paper \cite{cham_1} this ansatz was made in the following form
\begin{equation}\label{cham_7}
H=\exp\left[\frac{1-\gamma a^2}{\alpha a}\right],
\end{equation}
where $\gamma$ and $\alpha$ are positive constants. Choice of the ansatz is dictated by desire to reproduce well the observed expansion of the Universe. The system of equations is now closed, but the high degree of non- linearity forces us to make the additional simplifying assumption \cite{cham_1}
\begin{equation}\label{cham_8}
V=\beta\dot\varphi^2.
\end{equation}
The condition $\beta>1/2$ guarantees negative pressure of the chameleon field. In presence of the additonal conditions (\ref{cham_7}) and (\ref{cham_8}) we must make sure that together with the equations (\ref{cham_2}-\ref{cham_4}) they provide a consistent set of solutions. Such solutions were obtained in \cite{cham_1}
\begin{equation}\label{cham_9}
f=\frac{a^3H^3}{\rho_0(\beta-1/2)}\left[(2\beta+1)\left(\frac1{\alpha a}+\frac{\gamma a}\alpha\right)-3\right];
\end{equation}
\begin{equation}\label{cham_10}
\dot\varphi^2=\frac{H^2}{\beta-1/2}\left[-2\left(\frac1{\alpha a}+\frac{\gamma a}\alpha\right)+3\right];
\end{equation}
Having convinced ourself in existence of the consistent set of solutions, we can analyze behavior of the DP
consistent set of solutions.
\begin{equation}\label{cham_11}
q=\frac{d}{dt}\left(\frac1H\right)-1=\frac{d}{dt}e^{-F(a)}=e^{-F(a)}\left(-\frac{dF}{da}\dot a\right)=\frac{1+\gamma a^2}{\alpha a}-1;
\end{equation}
where $F(a)\equiv(1-\gamma a^2)/(\alpha a)$. We can see that if $\alpha>1+\gamma$, then at present ($a=1$) expansion of the Universe is accelerated ($q<0$).

\begin{figure}
\includegraphics[width=\textwidth]{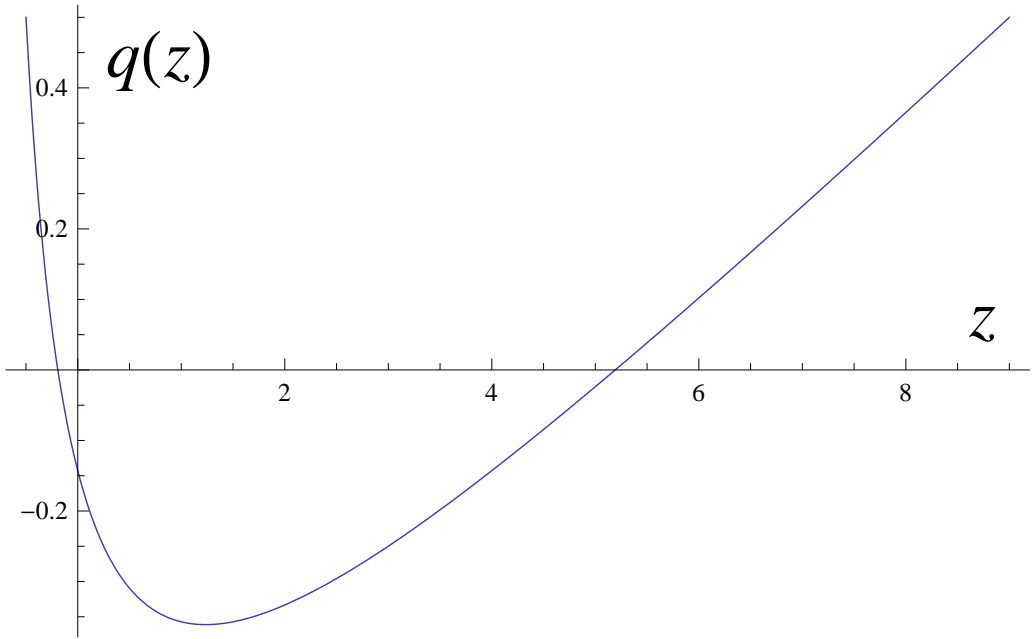}
\caption{\label{cham_f1} Plot of DP vs. redshift for $\alpha = 7$ and $\gamma = 5$.}
\end{figure}

A particular feature of the considered model is non-monotonic dependence of DP on the redshift (see Fig.\ref{cham_f1}).
\section{Phantom fields}
The enormous number of cosmological observations at our disposal shows that the parameter $w$  in the EoS for DE lies in a narrow range in the vicinity of $w =  - 1$. Above, we examined the region \( - 1 \le w <  - 1/3\). The lower boundary of the region, $w =  - 1$, corresponds to the cosmological constant, while the entire remaining interval can be realized with the aid of scalar fields with a canonical Lagrangian. We recall that the upper boundary \(w =  - 1/3\) is related to the necessity of providing the observed accelerating expansion of the Universe. Is it possible to go beyond the limits of this interval and realize $w <  - 1$? This is a complicated question for the energy component, of which we know so little. In GR, certain limits are conventionally imposed on possible values of the components of the energy-momentum tensor by energy conditions (see Part I, section 2.6). One of the simplest conditions of this type is \(\rho  + p \ge 0\). The physical motivation for this condition is to avoid the instability of the vacuum. Concerning the dynamics of the Universe, this condition requires the density of any admissible energy component not to increase with the expansion of the Universe. The cosmological constant, for which \({\dot \rho _\Lambda } = 0\), represents a limiting case. Taking into account our ignorance concerning the nature of DE, it is reasonable to ask ourselves: can this mysterious substance differ from the ''good'' sources of energy known to us and violate the condition \(\rho  + p \ge 0\)? Since DE must have a positive energy density (which is necessary for the Universe to be flat) and negative pressure (to provide the accelerated expansion of the Universe), this violation should lead to $w <  - 1.$

Some time ago, such a component, which was called the phantom energy, attracted the attention of physicists \cite{Nandra,PHANTOM,Caldwell}.

The action for a phantom field $\varphi $  minimally coupled to gravity differs from the canonical action for a scalar field by the sign of the kinetic term. The energy density and pressure of the phantom field
\begin{equation}
{\rho _\varphi } = {T_{00}} =  - \frac{1}{2}{\dot \varphi ^2} + V\left( \varphi  \right),\quad {p_\varphi } = {T_{ii}} =  - \frac{1}{2}{\dot \varphi ^2} - V\left( \varphi  \right), \label{23.1}
\end{equation}
and the EoS has the form
\begin{equation}
	{w_\varphi } = \frac{{{p_\varphi }}}{{{\rho _\varphi }}} = \frac{{{{\dot \varphi }^2} + 2V(\varphi )}}{{{{\dot \varphi }^2} - 2V(\varphi )}}.\label{23.2}
\end{equation}  	
If ${\dot \varphi ^2} < 2V(\varphi )$, then ${w_\varphi } <  - 1.$
As an example, we consider the case when the Universe contains only nonrelativistic matter  $(w = 0)$ and a phantom field $({w_\varphi } <  - 1).$ The densities of these components evolve independently: \[{\rho _m} \propto {a^{ - 3}}\] and \[{\rho _\varphi } \propto {a^{ - 3\left( {1 + {w_\varphi }} \right)}}.\] If matter domination terminates at a time moment ${t_m}$, the solution for the scale factor at $t > {t_m}$ is expressed as
\begin{equation}
a(t) = a({t_m}){\left[ { - {w_\varphi } + (1 + {w_\varphi })\left( {\frac{t}{{{t_m}}}} \right)} \right]^{\frac{2}{{3(1 + {w_\varphi })}}}}. 	 \label{23.3}
\end{equation}
Hence, it immediately follows that if ${w_\varphi } <  - 1,$ then, at the moment of time
\[{t_{BR}} = \frac{{{w_\varphi }}}{{(1 + {w_\varphi })}}{t_m}\]
the scale factor and a number of cosmological characteristics of the Universe (for instance, the scalar curvature and the energy density of the phantom field) become infinite. This catastrophe has been termed the Big Rip. The Big Rip is preceded by the so-called superaccelerating  expansion. A simple explanatory example helps to clarify the origin of the superacceleration regime. Let us consider the differential equation
\begin{equation}
\frac{{dx}}{{dt}} = A{x^2}. 	\label{23.4}
\end{equation}
If $A > 0$ , Eqn \eqref{23.4} realizes a positive feedback. The rapid increase of the function $x(t)$  leads to the Big Rip (the function blows up) in a finite time. Indeed, the general solution of Eqn \eqref{23.4} has the form
	\begin{equation}
x(t) =  - \frac{1}{{A(t + B)}},
	\label{23.5}
\end{equation}
where B is the integration constant. At $t =  - B$ , the Big Rip occurs.
It is easy to see that model  \eqref{23.4} represents a concrete version of the Friedmann equation for ${w_\varphi } <  - 1.$  Because ${\rho _\varphi } \propto {a^{ - 3(1 + {w_\varphi })}},$ the first Friedmann equation can be represented as
\begin{equation}
\dot a = A{a^{ - \frac{3}{2}\left( {1 + {w_\varphi }} \right) + 1}}.	\label{23.6}
 \end{equation}
For example, for ${w_\varphi } =  - 5/3,$ Eq. \eqref{23.6} coincides precisely with \eqref{23.4}.
Note that on the contrary to the phantom dominated case $\left( {1 + {w_\varphi } < 0} \right)$ in the quintessence dominated Universe $\left( {1 + {w_\varphi } > 0} \right),$ the cosmic expansion is singularity-free with the scale factor
\begin{equation}
a(t) = a\left( {{t_m}} \right){\left[ {1 + \left( {1 + {w_\varphi }} \right)\left( {t - {t_m}} \right)/{t_m}} \right]^{\frac{2}{{3\left( {1 + {w_\varphi }} \right)}}}}.
\end{equation}
Let us consider for instance a phantom scenario in the case of hybrid expansion law \cite{Weinberg}. In this case the anzatz
\begin{equation}
a(t) = {a_0}{\left( {\frac{t}{{{t_0}}}} \right)^\alpha }{e^{\beta \left( {\frac{t}{{{t_0}}} - 1} \right)}}
\end{equation}
must be slightly modified in order to acquire self consistency. It can be achieved by rescaling the time as $t \to {t_s} - t$, where ${t_s}$  is a some succinctly positive reference time.
The rescaling leads to
\begin{equation}
a(t) = {a_0}{\left( {\frac{{{t_s} - t}}{{{t_s} - {t_0}}}} \right)^\alpha }{e^{\beta \left( {\frac{{{t_s} - t}}{{{t_s} - {t_0}}} - 1} \right)}}	 \label{23.7}
\end{equation}
and
\begin{equation}
H = \frac{\alpha }{{{t_s} - t}} - \frac{\beta }{{{t_s} - {t_0}}},\quad \dot H = \frac{\alpha }{{{{\left( {{t_s} - t} \right)}^2}}},\quad q = \frac{{\alpha {{\left( {{t_s} - {t_0}} \right)}^2}}}{{{{\left[ {\beta \left( {{t_s} - t} \right) + \alpha {{\left( {{t_s} - {t_0}} \right)}^2}} \right]}^2}}} - 1.
\label{23.8}
\end{equation}
It then immediately follows that $\alpha  < 0$ leads to $q < 0$ (acceleration). Moreover, as $\dot H > 0$, then this case realizes the superacceleration $q <  - 1\;,\dot H =  - {H^2}(1 + q).$
Also the scale factor and Hubble parameter diverge as $t \to {t_s}$  and thus exposing the Universe to Big Rip.

It is also worth noting that in this models at late times the DP tends to the constant value
\begin{equation}
q(t)\rightarrow-\frac{1}{2}|1+3w|.
\end{equation}
The cosmological models with the superacceleration evidently lack big chance to be realized in our Universe, as it would make impossible the large-scale structures to appear unless too fine tuning of the model parameters is performed. The Universe evolution rather realizes a soft violation of the equality $w=-1$, but in such a way that does not (and will not) lead to decay of cosmological structures or to the Big Rip. We consider models of such type below.
\section{Crossing the Phantom Divide}
As was mentioned in the previous section, there exists certain observational evidence in favor of the fact that the EoS parameter takes the so-called fantom value $w<-1$ in the present time or in the recent past.
\begin{figure*}[t]
\includegraphics[width=\textwidth]{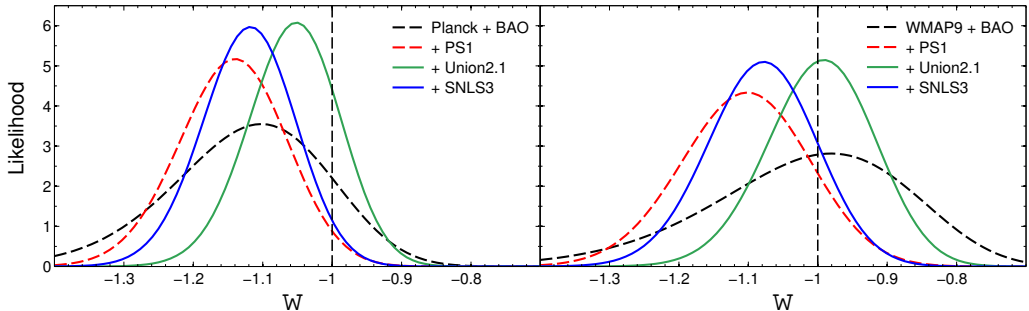}
\caption{Likelihood curves for a constant EoS $w$ in a flat universe, using Planck CMB data (left panel) and WMAP9 CMB data (right panel). In the figure are compare constraints from CMB + BAO data alone (dashed black) to those which additionally include SN Ia data from SNLS3 $(1\sigma)$  \cite{SNLS3} (blue), Union2.1 (green), or PS1 $(1.2\sigma)$ \cite{Pan-STARRS1}  (dashed red). All likelihoods are marginalized over other cosmological and nuisance parameters\cite{Huterer:2004ch}.}
\label{fig:wlike}
\end{figure*}
The best current measurement \cite{nakamura10} for the EoS (assumed constant) is
\begin{equation}
w=-1.04^{+0.09}_{-0.10}. \label{2}
\end{equation}
Figure \ref{fig:wlike} shows confidence intervals for the EoS parameter obtained with the help of different observational data.

Constraints on the DE EoS from recent SN Ia data complemented with distance measurements from the CMB and a compilation of BAO results. For the SNLS3 and PS1 SN data sets, the combined SN Ia + BAO + Planck data favor a phantom EoS where $w < -1$ at $\sim 1.9\sigma$ confidence (see Fig.~\ref{fig:wlike}), in good agreement with the corresponding results reported in the original SN Ia and CMB data set.

On the other hand, it is commonly known that the DE with phantom EoS violates the dominant energy condition which is a sufficient condition of the conservation theorem \cite{Hawking}, and therefore the vacuum in such models is unstable both on classical and quantum levels.

Actually, in the articles \cite{Hoffman,Cline} they are showing that the scalar fields with $w < -1$ are quantum-mechanically unstable concerning decay of the vacuum into gravitons and phantom particles with negative energy. This problem could be solved by introduction of the DE with time-depending EoS parameter.
If the DE could dynamically change its EoS  from a phantom like one to that with $w\geq-1$,
then this crossing would stave off the unacceptable particle production. As was noted in Ref.~\cite{Onemli,Onemli2}, quantum effects in locally de Sitter space could lead to the effective parameter $w<-1$.

A further fundamental physical question where this transition could play
a significant role is the cosmological singularity problem. If $w<-1$
in an expanding Friedmann Universe, then the positive energy density
of such phantom matter commonly becomes infinite in finite time,
dominating over all other forms of matter and, consequently, leads to the late-time
singularity called the ``Big Rip'' \cite{Big_rip}. The models where EoS parameter grows during the evolution and the DE transits from the regime with $w<-1$ to that with $w\geq-1$
naturally prevent this late-time singularity.
Here it is worthwhile to mention that for certain potentials and initial
conditions the phantom scalar fields can escape this singularity by
evolving to a late-time asymptotic which is the de Sitter solution
with $w=-1$. Moreover, it was argued that the
quantum effects can prevent the developing of the {}``big-rip''
singularity as well.
In the present subsection we consider a model of flat Universe filled with three components: the DE $\rho_{de}=\rho_1+\rho_2$ with partial energy densities $\rho_1$ and $\rho_2$, and non-relativistic matter $\rho_w$
\begin{equation}
H^2=\frac{1}{3M_p^2}(\rho_1+\rho_2+\rho_w)
 \label{fried}
 \end{equation}
The EoS for the DE reads
\begin{equation}
 \label{rho1}
  \frac{\rho_1}{\rho_0}=\sqrt{A+B(1+z)^6},
 \end{equation}
 \begin{equation}
   \frac{\rho_2}{\rho_0}=-\sqrt{C+D(1+z)^6}.
  \label{rho2}
 \end{equation}
Here $\rho_1$  denotes the density of type I Chaplygin gas and $\rho_2$ represents the density of type II Chaplygin gas.

The continuity equation then is
 \begin{equation}
 {d\rho_{de}}=3(\rho_{de}+p_{de})d\ln(1+z).
 \label{conti}
 \end{equation}
 \begin{figure}
 \centering
 \includegraphics[width=0.7\textwidth]{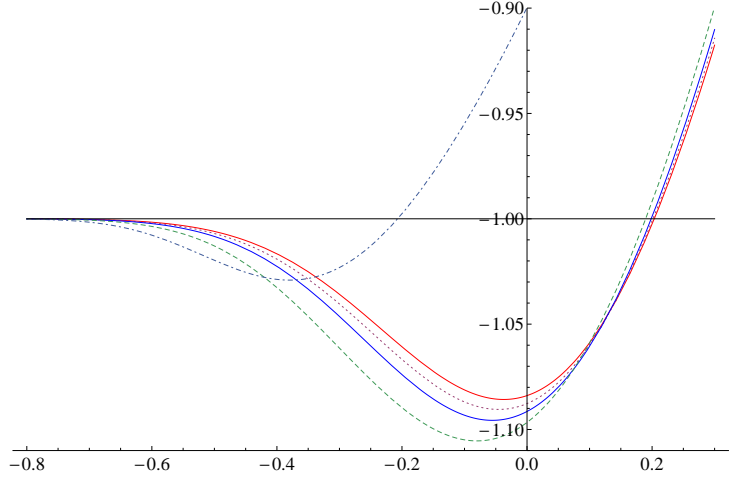}
 \caption{$w_{de}(z)$ as a function of $z$. On all of the 5
 orbits $B=0.3,~D=0.15$ with different $A$ and $C$.
 From the above to the below $A=1.104,C=0.048;~~A=1.059,C=0.039;~~A=1.01,~C=0.03;~~
 A=0.9106,C=0.015;~~A=0.8267,C=0.006$.}
 \label{w_cross}
 \end{figure}
\begin{figure}
\centering
\includegraphics[width=0.7\textwidth]{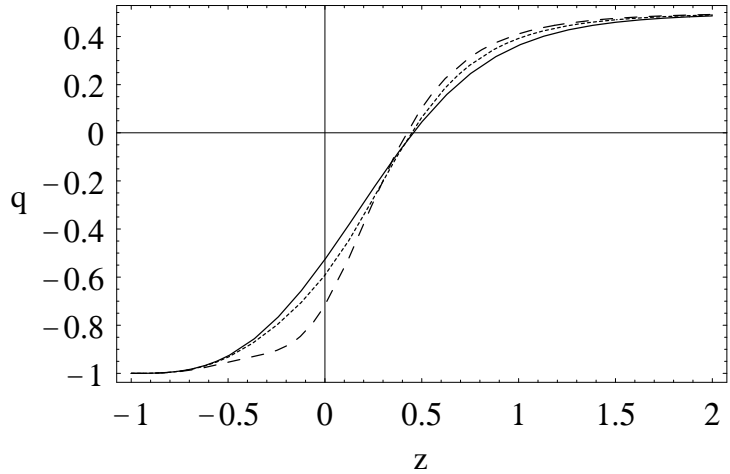}
\caption{DP $q$ as a function of $z$. On all of
the curves orbits $B=0.3,~D=0.15,~ \Omega_m=0.28$ with different $A$
and $C$. For the solid curve $A=5.92,~C=3.00$, the short dashed
curved $A=1.63,~C=0.300$, while for the long dashed curve
$A=1.01,~C=0.0300$. }
 \label{dece}
 \end{figure}
From the continuity equation (\ref{conti}), we arrive at
    \begin{equation}
  w_{de}=\frac{p_{de}}{\rho_{de}}=-1-\frac{1}{3}\frac{d \ln \rho_{de}}{d \ln
  (1+z)},
  \end{equation}
which means that in an expanding Universe
if $\rho_{de}$ decreases and then increases
w.r.t. the redshift, or increases and then
decreases, then we conclude that EoS of DE crosses phantom
divide. Here,
\begin{equation}
  \frac{d\rho_{de}}{dz}=-3[B\rho_1^{-1} (1+z)^5+D\rho_2^{-1}
  (1+z)^5].
\end{equation}
Therefore one obtains the EoS parameter as
\begin{equation}
  w_{de}=\frac{p_{de}}{\rho_{de}}=-1+(1+z)^6[B\rho_1^{-1} +D\rho_2^{-1}].
\end{equation}
The equation $w_{de}(z)=-1$ has a solution. It is easy to see that at $z \to -1, \, w_{de}(z)=-1$
\begin{equation}
  z_{tr}=-1+\sqrt[6]{\frac{AD^2-B^2C}{BD(B-D)}}.
\end{equation}
If this expression is treated as a small addition to $-1$ then one can conclude that the regime $w_{de}\simeq-1$ is realized on late stages of evolution of Universe.
We see that if we carefully tune $B$ and $D$, the EOS of dark
energy may cross $-1$. Figure \ref{w_cross} shows some concrete
examples for which Chaplygin crosses the phantom divide at about $z=0.2$.

We now turn to analysis of the DP in this model which is the most important parameter from the observational point of view. It is known that $q$ carries the total
  effects of cosmic fluids. Here $q$ reads
 \begin{equation}
  q=-\frac{\ddot{a}a}{\dot{a}^2}=\frac{1}{2}\left[1-3(\frac{A}{\rho_1}+\frac{C}{\rho_2})
  \frac{1}{\rho_1+\rho_2+\rho_m}\right],
   \end{equation}
where $\rho_1$,$\rho_2$ are defined in the expression \eqref{rho1} and \eqref{rho2} respectively, and $\rho_w$ is the DE with the EoS parameter $w:$ $\rho_w  = \rho_{w0} (1+z)^{3(1+w)} $.
 \begin{equation}
q (z=0) =\frac{1}{2} -\frac{3}{2}\frac{\left(\frac{A}{\sqrt{A+B}}-\frac{C}{\sqrt{C+D}}\right)}{\sqrt{A+B}-\sqrt{C+D}+1}.
  \end{equation}
In the infinitely far future, i.e. at $z\to -1$, we obtain the de Sitter Universe with $q\to -1$, and the divergence problem is absent.
\section{Quintom Models}
Analysis of properties of the DE from recent observations mildly favor models with $w$ crossing $-1$ in the recent past. However, as was shown above, neither quintessence nor phantom can realize this transition. But this transition can be easily realized in the Universe where role of the DE is played by quintessence and phantom fields simultaneously. This kind of DE is termed as quintom \cite{Feng_Wang_Zhang}. Besides that, this model for certain parameter values describes the results of observations better than most of other models with $w \geq -1$. Moreover, the quintom DE models provide the so-called tracking field regime and therefore facilitate solution of the coincidence problem, in distinction to the phantom DE which suffers from this problem particularly heavy \cite{Li_Zhang}.

In the present subsection we give theoretical basis of the quintom DE and also describe some models which reflect its characteristic features.

As we mentioned before, the quintom model was initially suggested in the paper \cite{Feng_Wang_Zhang}, where the authors combined the Quintessence and Phantom components
together. The action for such a DE reads:
 \begin{equation}
 {\cal
}S_{Quintom}=\int
d^4x\sqrt{-g}\left\{\frac{1}{2}\nabla_\mu\varphi_1\nabla^\mu\varphi_1-\frac{1}{2}\nabla_\mu\varphi_2\nabla^\mu\varphi_2-V(\varphi_1,\varphi_2)\right\},
 \end{equation}
where $\varphi_1$ and $\varphi_2$ are Quintessence and Phantom components,
respectively.

The EoS for such compound DE can provide even multiple crossings of the phantom divide due to the combined effects of the two components. In general, the corresponding effective potential can have arbitrary form, and the fields can have a background interaction as well as a direct one via the energy exchange.

In the original paper \cite{Feng_Wang_Zhang} the authors consider  a simple model with the background interaction:
\begin{equation}
V(\varphi_1,\varphi_2)=V_0(e^{-\frac{\lambda\varphi_1}{m_{pl}}}+e^{-\frac{\lambda\varphi_2}{m_pl}})
\end{equation}
where $m_{pl}$ is the Planck mass and $\lambda$ is a
dimensionless constant.

As an example of the Universe with the quintom field we cite the model considered in paper \cite{Saridakis_Quintom}.

Let us consider the simple quintom cosmological model in a flat spacetime. The action in such a Universe consists of a
canonical $\varphi$ and a phantom $\sigma$ fields:
\begin{equation}
S=\int d^{4}x \sqrt{-g} \left[\frac{1}{2} R
-\frac{1}{2}g^{\mu\nu}\partial_{\mu}\varphi\partial_{\nu}\varphi+V_\sigma(\sigma)+
\frac{1}{2}g^{\mu\nu}\partial_{\mu}\sigma\partial_{\nu}\sigma+V_\sigma(\sigma)
+\cal{L}_\text{m}\right], \label{actionquint}
\end{equation}
where we have set $8\pi G=1$. Here $V_\varphi(\varphi)$ is the quintessence, $V_\sigma(\sigma)$ is the phantom field potentials,
 while the term $\cal{L}_\text{m}$ corresponds to the Lagrangian of the (dark) matter content of the
Universe, considered as dust. Radiation plays significant role on early stages of evolution of the Universe, but we consider the later epoch when it can be neglected for the sake of simplicity.

The Friedmann equations take on the form:
\begin{align}\label{FR1}H^{2}&=\frac{1}{3}\Big(\rho_{m}+\rho_{\varphi}+\rho_{\sigma}\Big),\\
\label{FR2}\dot{H}&=-\frac{1}{2}\Big(\rho_{m}+\rho_{\varphi}+p_{\varphi}+\rho_{\sigma}+p_{\sigma}\Big),
\end{align}
and the conservation equations for the quintessence and the phantom
fields are:
\begin{eqnarray}\label{eom}
&&\dot{\rho}_\varphi+3H(\rho_\varphi+p_\varphi)=0,\\
&&\dot{\rho}_\sigma+3H(\rho_\sigma+p_\sigma)=0,
\end{eqnarray}
 In these expressions,  $p_\varphi$ and $\rho_{\varphi}$ are respectively the
pressure and density of the quintessence, while  $p_\sigma$ and
$\rho_{\sigma}$ are the corresponding quantities for the phantom
field. Their energy densities and pressures are known to equal
\begin{eqnarray}\label{rhophiq}
 \rho_{\varphi}&=& \frac{1}{2}\dot{\varphi}^{2} + V_\varphi(\varphi)\\
 p_{\varphi}&=&  \frac{1}{2}\dot{\varphi}^{2} - V_\varphi(\varphi),\\
\label{rhosigma} \rho_{\sigma}&=& -\frac{1}{2}\dot{\sigma}^{2} + V_\sigma(\sigma),\\
 p_{\sigma}&=& - \frac{1}{2}\dot{\sigma}^{2} - V_\sigma(\sigma).\label{psigma}
\end{eqnarray}
Using the above given equations one can obtain the evolution equations for the quintessence and phantom energy:
\begin{eqnarray}
\label{phiddot} &&\ddot{\varphi}+3H\dot{\varphi}+\frac{\partial
V_\varphi(\varphi)}{\partial\varphi}=0\\
\label{sigmaddot}
 && \ddot{\sigma}+3H\dot{\sigma}-\frac{\partial
V_\sigma(\sigma)}{\partial\sigma}=0.
\end{eqnarray}
Using the exact solution $\rho_m=\rho_{m0}/a^3$ for the DM density, where
$\rho_{m0}$ is its present value, we obtain the closed system of equations.

In the quintom scenario, the DE field appears as a superposition of the quintessence $\varphi$ and phantom field $\sigma$:
$$\rho_{DE}\equiv \rho_{\varphi}+ \rho_{\sigma}$$
$$p_{DE}\equiv p_{\varphi}+ p_{\sigma}.$$
 Therefore, the DE EoS parameter takes on the form
\begin{equation}
\label{wdef}
w\equiv\frac{p_{DE}}{\rho_{DE}}=\frac{p_\varphi+p_\sigma}{\rho_\varphi+\rho_\sigma}.
\end{equation}
Below we follow the paper \cite{Saridakis_Quintom} to consider the quintom evolution in power-law
potentials:
\begin{eqnarray} \label{potphi}
&&V_\varphi(\varphi) = \kappa_\varphi\varphi^{-\alpha_\varphi}\\
&&V_\sigma(\sigma) = \kappa_\sigma\sigma^{-\alpha_\sigma},
 \label{pots}
\end{eqnarray}
where $\kappa_\varphi$ and $\kappa_\sigma$ are constants with units
$m^{4+\alpha_\varphi}$ and $m^{4+\alpha_\sigma}$ respectively.
As was shown in the paper \cite{Saridakis_Quintom} the weak energy condition requires $\alpha_\varphi$ to be positive and $\alpha_\sigma$ to be negative
and bounded, that is potential (\ref{potphi}) will be an inverse
power-law one while (\ref{pots}) will be of a normal power-law
form.

It is convenient to rewrite the system \eqref{phiddot}-\eqref{sigmaddot} in terms of the scale factoe
\begin{eqnarray}
 && x_\varphi =
{\rho_\varphi + p_\varphi\over 2(\rho_m + \rho_\varphi)} = {{1\over 2}
\dot\varphi^{2}\over 3H^2} = {1\over 6}{\Big(a{d\varphi\over
da}\Big)}^2,\\
 && x_\sigma=
{\rho_\sigma + p_\sigma\over 2(\rho_m + \rho_\sigma)} = {-{1\over 2}
\dot\sigma^{2}\over 3H^2} = -{1\over 6}{\Big(a{d\sigma\over da}\Big)}^2,
\end{eqnarray}
and thus (\ref{rhophiq}),(\ref{rhosigma}) can be written as:
 \begin{eqnarray}
 \label{phidotx}
\frac{1}{2} \dot\varphi^2 = {x_\varphi\over 1-x_\varphi}[\rho_m + V_\varphi(\varphi)],\\
-\frac{1}{2} \dot\sigma^2 = {x_\sigma\over 1-x_\sigma}[\rho_m + V_\sigma(\sigma)].
\label{sigmadotx}
\end{eqnarray}
 Therefore, we can
simply write:
\begin{eqnarray}
 &&\rho_{\varphi} = \frac{x_\varphi\rho_m + V_\varphi(\varphi)}{1-x_\varphi}\label{rhox}\\
 &&\rho_{\sigma} = \frac{x_\sigma\rho_m + V_\sigma(\sigma)}{1-x_\sigma}\label{rhoxs}\\
 &&p_{\varphi}=\frac{x_\varphi\rho_m - V_\varphi(\varphi)(1-2x_\varphi)}{1-x_\varphi}\label{px}\\
 &&p_{\sigma}=\frac{x_\sigma\rho_m - V_\sigma(\sigma)(1-2x_\sigma)}{1-x_\sigma}\label{pxs},
\end{eqnarray}
and thus for the DE EoS parameter we
obtain:
\begin{equation}
 w = \frac{\frac{x_\varphi\rho_m - V_\varphi(\varphi)(1-2x_\varphi)}{1-x_\varphi}+\frac{x_\sigma\rho_m - V_\sigma(\sigma)(1-2x_\sigma)}{1-x_\sigma}}
 {\frac{x_\varphi\rho_m + V_\varphi(\varphi)}{1-x_\varphi}+\frac{x_\sigma\rho_m + V_\sigma(\sigma)}{1-x_\sigma}}\label{wx}.
\end{equation}
Similarly, the first Friedman equation can be written as
\begin{equation}
3H^2 = \frac{\rho_m + V_\varphi(\varphi)+V_\sigma(\sigma)}{1-x_\varphi-x_\sigma}\label{Hx}.
\end{equation}
 Finally, using
expressions (\ref{rhox})-(\ref{pxs}) and (\ref{Hx}) the field
evolution equations (\ref{phiddot}) and (\ref{sigmaddot}) become
\begin{align}
\nonumber a^{2}{\varphi}'' &+ {a{\varphi}'\over 2}\left(5-3x_\varphi-3x_\sigma\right) +\\
&+{3(1-x_\varphi-x_\sigma)\over \rho_m + V_\varphi(\varphi)+V_\sigma(\sigma)}\Big\{{a{\varphi}'[V_\varphi(\varphi)+V_\sigma(\sigma)]\over
2}+ {dV_\varphi(\varphi)\over d\varphi}\Big\} = 0 \label{eomxp}\\
\nonumber a^{2}{\sigma}'' &+ {a{\sigma}'\over 2}\left(5-3x_\varphi-3x_\sigma\right) +\\
&+{3(1-x_\varphi-x_\sigma)\over \rho_m + V_\varphi(\varphi)+V_\sigma(\sigma)}\Big\{{a{\sigma}'[V_\varphi(\varphi)+V_\sigma(\sigma)]\over
2}- {d V_\sigma(\sigma)\over d\sigma}\Big\} = 0,  \label{eomxs}
\end{align}
with prime denoting the derivative w.r.t. $a$. These
equations are exact and account for the complete dynamics of the
quintom scenario.

\begin{figure}
\centering
\includegraphics[width=0.48\textwidth]{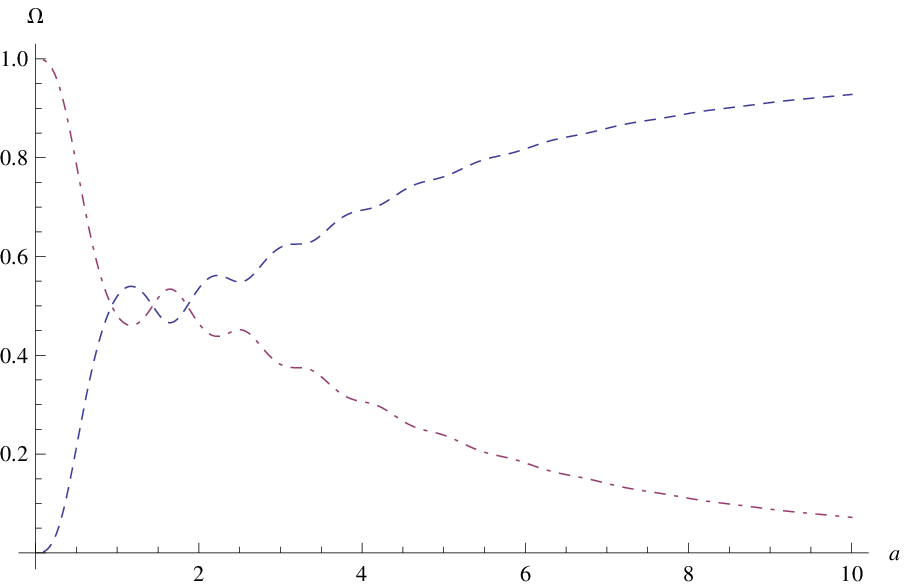}
\includegraphics[width=0.48\textwidth]{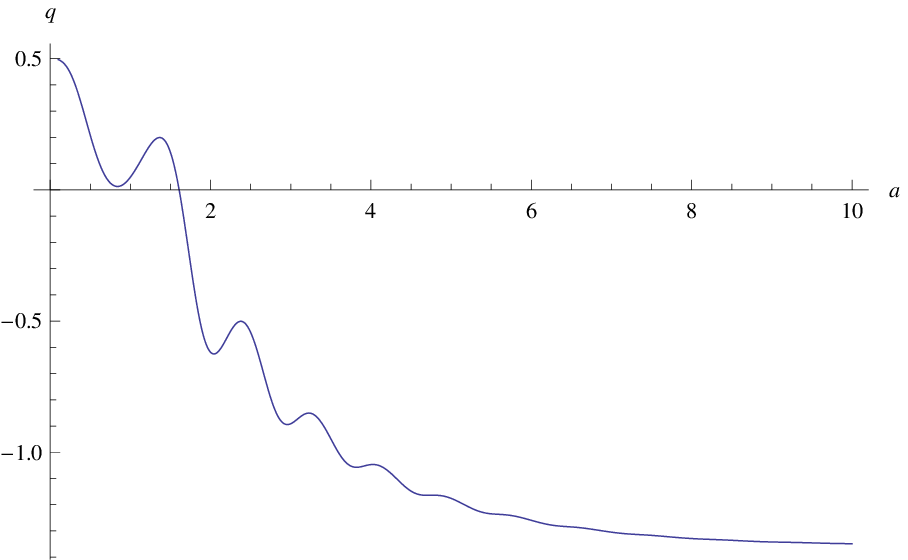}
\caption{Relative densities $\Omega_m, \Omega_{de}$ and the DP $q$ as a function of $a$. All curves correspond to $k_1= 2,~k_2=0.5, \alpha = -2, \beta = 3~ \rho_m=1$  }
 \label{dece_q}
 \end{figure}

Unfortunately there is no exact solution for the system (\ref{eomxp},\ref{eomxs}), however it is possible to obtain some exact results (see \cite{Saridakis_Quintom} for details) in the limit of non-relativistic matter domination, i.e.
$\rho_\varphi,|p_\varphi|\ll\rho_m$ and $\rho_\sigma,|p_\sigma|\ll\rho_m,$ which is relevant only for early Universe, as at the present time the DE density is known to be twice greater than that of the DM. We give the results of numerical integration of the system (\ref{eomxp},\ref{eomxs}), and according to the context of our review, instead of the EoS (as it was in \cite{Saridakis_Quintom})  we make accent on the DP $q,$ taking into account that the relative density has manifestly non-monotonic dependence (see Figure \ref{dece_q}). This is non-trivial distinction, at least in the range $ 1\lesssim a\gtrsim 1,$ for the chosen values of the model parameters.

DE domination on early stages of evolution of the Universe is characteristic for all DE models. It is the case also for the considered model: in the early Universe we get $q(a\to 0)\to 1/2$. While the quintom contribution in the energy balance grows, this simple picture gets destructed and the dependence $q(a)$ acquires clearly non-monotonic character. The first local minimum appears at $a\approx 0.862, q\approx 0.0135$, then follows a short phase of decelerated expansion, and the next local minimum at $a\approx 1.39, q\approx 0.2.$ After several pulsations of relative densities (and respective oscillations of the DP with general trend to accelerated expansion), relative contribution of the DE starts to dominate, and the Universe enters to the stage of accelerated (and soon the super-accelerated) expansion. It worth noting that this model is free of the Big Rip in spite of the fact that during a long period of evolution of the Universe the equivalent EoS parameter lies within the phantom zone, and the Universe is in the super-acceleration regime. The minimum value of the DP for the given values of the model parameters equals to $q_{\min}\approx-1.35,$ after which it starts asymptotic approach to the value $q(a\to \infty)\to-1.$
\section{Interaction in the dark sector}
SCM considers DM and DE as independent components
of energy budget of the Universe. DE in the SCM is postulated in form of
the cosmological constant, introduced as early as by Albert Einstein in
order to make possible the creation of stationary Universe model.
The assumed absence of interaction between the two components means
that the energy densities for each component obey independent
conservation equations
\[{{\dot{\rho }}_{i}}+3H\left( {{\rho }_{i}}+{{p}_{i}} \right)=0.\]
Coupling between the two components leads to modified evolution of the
Universe. In particular, the energy density for non-relativistic
component will not evolve according the law ${{a}^{-3}},$ and the
DE density (in form of the cosmological constant) will not
remain constant any more. From one hand, such modification of the
theory opens for us new possibilities for solution of principal
problems in cosmology. So, for example, solution of the coincidence
problem\footnote{the problem is that during all the Universe history
the two densities decay by different laws, so it is necessary to
impose very strict limitations on their values in early Universe in
order to make them to be of comparable order nowadays} reduces to
appropriate choice of the interaction parameters \cite{InteractionQ10,InteractionQ8}, which can satisfy the condition \[\frac{{{\Omega }_{de}}}{{{\Omega
}_{dm}}}={\mathrm O}(1)\] in the present time consistently with the
condition $\ddot{a}>0.$ From the other hand, introduction of the
interaction will modify the relations between the observable
parameters. In particular, the modification will affect the
fundamental relation between the photometric distance and the
redshift, which the evidence of accelerated expansion of the Universe is
mainly based on. It imposes strict limitations both on the form of
the interaction and its parameters.

Taking into account that the DM and the DE represent
dominant components in the Universe, from point of view of the
field theory it is naturally to consider interaction between them
\cite{InteractionQ1,InteractionQ2,InteractionQ3,InteractionQ4,InteractionQ5,InteractionQ6,InteractionQ7}.
Non-minimal coupling in the dark sector can considerably affect history of the
cosmological expansion of the Universe and evolution of the density
fluctuations, thus modifying the rate of the cosmological structure
growth. Different and independent data of many observations, such as
Wilkinson Microwave Anisotropy Probe, SNe Ia and BAO, were specially
analyzed in order to investigate the limitations on the intensity
and form of the interaction in the dark sector. Some researchers
also suggest \cite{BaldiClusters}, that dynamical equilibrium of the
collapsing structures, such as clusters, will essentially depend on
form and sign of the interaction between the DM and the DE.
\subsection{Model of Universe with time dependent cosmological constant}
The simplest example of the model with interacting DM and DE is the
cosmological model with decaying vacuum. Actually the
$\Lambda$(t)CDM cosmology represents one of the cases where the EoS parameter $w$ for DE equals to $-1.$

This model is based on the assumption that the DE is
nothing that physical vacuum, and energy density of the latter is
calculated on the curved space background with subtraction of the
renormalized energy density of physical vacuum in the flat space
\cite{0711.2686}. The resulting effective energy density of physical
vacuum depends on space-time curvature and decays from high initial
values in the early Universe (at strong curvature) to almost zero
magnitude in the present time.

Due to the Bianchi identity, the decay of vacuum must be accompanied
by creation or mass increase of the DM particles, which is
common property of the decaying vacuum models, or in more general
case, for the models with interacting DM and DE.

Now let us give the formulation of the $\Lambda$(t)CDM model. In the
present section we will use the system of units where the reduced
Planck mass equals to unity: ${{M}_{Pl}}={{\left( 8\pi G
\right)}^{-1/2}}=1.$

For the case of flat Universe described by FLRW-metric, the first
Friedmann and the conservation equation can be
presented in the form
\[ \rho _{tot}  = 3H^2,~~\dot \rho _{tot}  + 3H(\rho _{tot}  + p_{tot} ) = 0,\]
where $\rho _{tot}  = \rho _m  + \rho _\Lambda,$  and taking into
account that $p_m  = 0$ and $p_\Lambda   = - \rho_\Lambda,$ the
conservation equation takes on the form:
\begin{equation}\label{3.1.1}
    \dot \rho _m  + 3H\rho _m  =  - \dot \rho _\Lambda.
\end{equation}
The right hand side of the latter contains an additional term which
plays role of a source --- it is the decaying cosmological constant.
The problem with such approach is in fact that we end up with the
same number of equations, but acquire additional unknown function
$\Lambda(t).$ The above described approach to define the physical
vacuum density, though intuitively clear, faces certain principal
difficulties \cite{BertolamiLdec}. That is why a phenomenological
approach prevails in the literature. Below we present a simple but
exactly solvable model. Consider the case when $\Lambda$ depends on
time as \cite{Carneiro}:
\begin{equation}\label{3.1.2}
    \Lambda  = \sigma H.
\end{equation}
It is interesting to note that with the choice $\sigma \approx
m_{\pi }^{3}$ (${{m}_{\pi }}$ is the energy scale of the QCD vacuum
condensation) the relation (\ref{3.1.2}) gives the value of
$\Lambda$ which is very close to the observed one.

For the considered components the first Friedmann equation takes on the
form
\begin{equation}\label{lamd_4}
{{\rho }_{\gamma }}+\Lambda =3{{H}^{2}}.
\end{equation}
The equations (\ref{3.1.1}) and (\ref{lamd_4}) together with the
conservation equation
\[{{p}_{\gamma }}=w{{\rho }_{\gamma }}\equiv \left( \gamma -1 \right){{\rho }_{\gamma }},\]
and the decay law (\ref{3.1.1}) for the cosmological constant
completely determine time evolution of the scale factor. Combining
those equations, one obtains the evolution equation in the following
form
\[2\dot{H}+3\gamma {{H}^{2}}-\sigma \gamma H=0.\]
Under the condition $H>0$ the solution for the latter equation reads
\[a(t)=C{{\left[ \exp \left( \sigma \gamma t/2 \right)-1 \right]}^{\frac{2}{3\gamma }}},\]
where $C$ is one of the two integration constants (\cite{Bolotin_et_al} for details).

This expression can be used for calculation of the DP
\begin{equation}\label{lamd_22}
q(z)=\frac{\frac{3}{2}{{\Omega }_{m0}}{{(1+z)}^{3/2}}}{1-{{\Omega }_{m0}}+{{\Omega }_{m0}}{{(1+z)}^{3/2}}}-1.
\end{equation}

One thus finds for the current value of the DP
    \[q(z=0)=\frac{3}{2}{{\Omega }_{m0}}-1.\]
Therefore the accelerated expansion of the Universe occurs under the following condition
\[{{\Omega }_{m0}}<\frac{2}{3},\]
This condition is in fact satisfied for the observed value ${{\Omega
}_{m0}}\approx 0.23.$ From (\ref{lamd_22}) it follows that the
transition from the decelerated expansion to the accelerated one
took place at
    \[z^*={{\left[ 2\frac{1-{{\Omega }_{m0}}}{{{\Omega }_{m0}}} \right]}^{2/3}}-1.\]
This value $\left( z^*\approx 1.2 \right)$ exceeds (being however of
the same order) the corresponding value for the SCM
$\left( z^*\approx 0.75 \right),$ which is the result of the matter
production in the vacuum decay process.

Therefore the model based on the time-dependent cosmological
constant, where vacuum density linearly depends on the Hubble parameter,
appears to be quite competitive. It sufficiently accurately
reproduces the ''canonic'' results, relevant both to the radiation
dominated and matter dominated eras. The present Universe expansion
is accelerated according to the model under consideration.
Verification of the model with the latest observational data
obtained for SN1a leads to the results (e.g., $0.27\le {{\Omega
}_{m0}}\le 0.37 $), that agree very well with the currently accepted
estimates for the parameters of the accelerated expansion of the Universe.
\subsection{Varying $\Lambda$ as a function of the scale factor.}
Let us now consider a flat two-component Universe filled by matter with the EoS $p=w\rho$ (or $p=(\gamma-1)\rho$ in terms of $\gamma = w+1$) and cosmological constant with the following scale factor dependence
\begin{equation}
\Lambda = {\cal B} \, a^{-m}.
\label{Bam}
\end{equation}
The conservation equation and the first Friedman equation can be used to obtain the ODE
 \begin{equation}
\frac{d}{da} \left( \rho a^{3\gamma} \right) = \left( \frac{m{\cal B}}{8\pi G}
                                               \right) a^{3\gamma-(m+1)} .
\end{equation}
Its integration leads to the energy density of matter as a function of the scale factor
\begin{equation}
\rho (a) = \rho_0 a^{-3\gamma} f(a) ,
\label{NewRho}
\end{equation}
where as usual $a(t_0)=a_0=1,\rho(a_0)=\rho_0$, and the function $f(a)$ takes on the form
\begin{equation}
f(a)  \equiv  1 + \kappa_0 \times \left\{ \begin{array}{ll}
                    \frac{ {\textstyle m(a^{3\gamma-m}-1)} }
                         { {\textstyle 3\gamma-m} }
                          & \mbox{ if } m \neq 3\gamma, \\
                    3\gamma \ln(a)
                          & \mbox{ if } m=3\gamma,
                    \end{array} \right.
                    \label{fDefn}
\end{equation}
where $\kappa_0  \equiv  {\cal B}/8\pi G \rho_0$. If $m=0$ then $f(a)=1$ and the equation (\ref{NewRho}) recovers the usual result $\rho\sim a^{-3}$ for the non-relativistic matter ($\gamma=1$) and $a^{-4}$ for the radiation ($\gamma=4/3$). The new parameter $\kappa_0$ can be determined by the decay law (\ref{Bam}) and astronomical observations imply ${\cal B} = \Lambda_0 = 3H_0^2\lambda_0$, then $\kappa_0 = \lambda_0/\Omega_0$.

\begin{figure}[h]
\centering \label{Lambda_gamma1-43m2}
\includegraphics[width=0.47\textwidth]{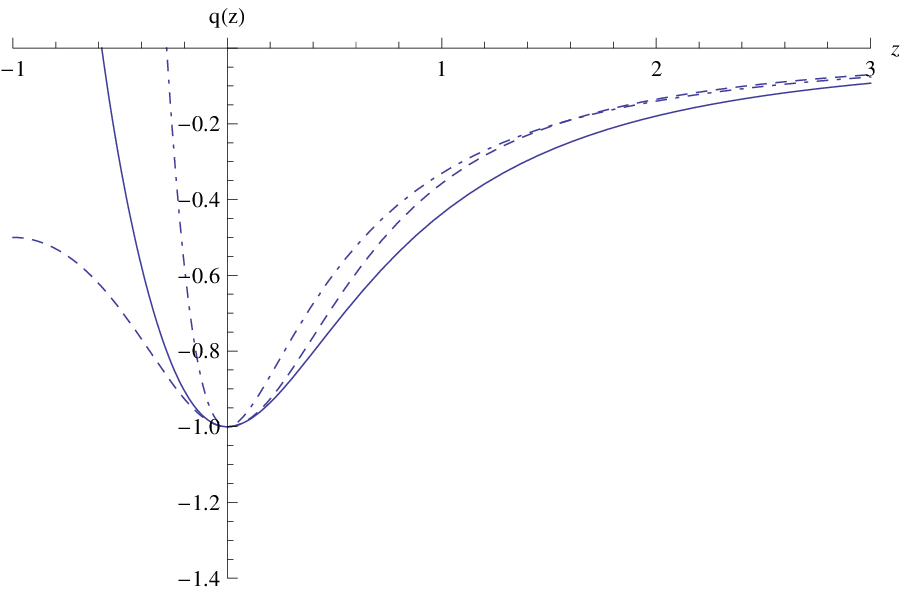}
\includegraphics[width=0.47\textwidth]{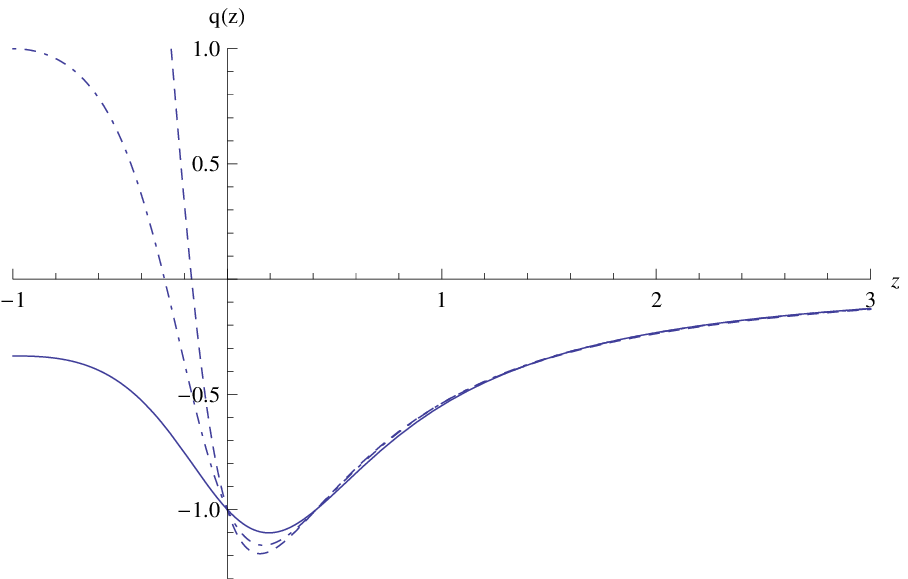}
\caption{The $q(z)$ dependence for the model (\ref{Bam}). On the left: for $\Omega_0 = 1,~\gamma=1, \lambda_0 = 1/2,$ and $m=1$ (dashed), $m=2$ (solid), $m=3$ (dash-dot). On the right: for $\gamma=4/3$ and $m=1$ (solid); $m=2$ (dash-dot), $m=3$ (dashed).}
\label{figSpectra}
\end{figure}

Substituting (\ref{Bam}) and (\ref{NewRho}) into the Friedman equation one obtains
\begin{equation}
\frac{da}{d\tau} = a \left[ \Omega_0 a^{-3\gamma} f(a) -
                   (\Omega_0+\lambda_0-1) a^{-2} +
                   \lambda_0 a^{-m} \right]^{1/2} ,
\label{Expansion}
\end{equation}
where the function $f(a)$ is determined in \eqref{fDefn}.
The second Friedman equation can be rewritten in the form
\begin{equation}
\frac{d^{\, 2}a}{d\tau^2} = \left( 1 - \frac{3\gamma}{2} \right)
                            \Omega_0 a^{1-3\gamma} f(a) +
                            \lambda_0 a^{1-m} .
\label{Deceleration}
\end{equation}
Substituting this derivatives into the definition of the DP, one obtains
\begin{equation}
q\equiv -\frac{\ddot{a}a}{\dot{a}^2}=-\frac{\left( 1 - \frac{3\gamma}{2} \right)\Omega_0 a^{1-3\gamma} f(a) +                            \lambda_0 a^{1-m}}{a\left[ \Omega_0 a^{-3\gamma} f(a) -(\Omega_0+\lambda_0-1) a^{-2} +\lambda_0 a^{-m} \right]}
\end{equation}
Depending on values of the model parameters it is possible to obtain qualitatively different types of evolution of the Universe (see Figure \ref{Lambda_gamma1-43m2}). In the case $m>1$, the Universe enters into the decelerated expansion stage starting from the $z = 0$, and in the infinitely distant future expansion of the Universe stops \[q(m>1,z\to -1)\to \infty.\] For $m = 1$, we obtain accelerated expansion of the Universe in the infinitely distant future \[q(m=1,z\to -1) \to -\frac12.\] A distinguishing feature is the finite value of the DP at the distance future: for $m=1;3$ we have $ q(z\to -1) \to -1 ,$  for $m=3$ we have $ q(z\to -1) \to 5 $ and $ q(z\to -1) \to 1 $ for $m=2.$
We note that, the appropriate choice of the parameter values, essentially of the parameter $\lambda_0$ which is responsible for the decay intensity, would make the DP to take only finite values (at least until $m\leq 3$).

Thus, in this simple model there is a large freedom for the fitting. As might be expected, with the expansion of the Universe the contribution of the cosmological constant is reduced and the rate of the accelerated expansion of the Universe also changes, which is reflected by changes of the DP. There are three different types of the distant future: eternal slowing down, eternal speedup or the so-called stopping of the Universe, i.e. expansion of the Universe with ever increasing value of the DP: \[q(z\to -1) \to \infty.\]
\subsection{Interacting DM and DE}
One of the most interesting properties of the DM is its
possible interaction with the DE. Although the most realistic
models (the SCM in particular) postulate that the DM and DE are uncoupled, there is no serious evidence to consider such assumption as a principle. In the present time many active
researches are provided to investigate the possible consequences of
such interaction \cite{InteractionQ1,InteractionQ2,InteractionQ3,InteractionQ4,InteractionQ5,InteractionQ6,InteractionQ7,
InteractionQ10,InteractionQ8,InteractionQ9, 0801.1565,ZKGuo,
BaldiClusters, Chimento, Lima}. As we have mentioned above, the
interaction of the dark components can at least soften some sharp
cosmological problems, such as the coincidence problem for example.
The DE density is approximately three times as high as the
DM one. Such coincidence could be explained if DM
was somehow sensitive to DE.

Remark that the possibility of interaction between the DE
in form of scalar field and the DM lies in the basis of the warm
inflation model. Unlike the cold inflation scenario, the former does
not assume that the scalar field in the inflation period is isolated
(uncoupled) from other fields. That is why instead of the overcooled
Universe in the inflation period, certain quantity of radiation is always conserved,
which is sufficient to manifest itself in the post-inflation dynamics.

Interaction of the scalar field quantum, responsible for the
inflation start -- the inflaton -- with other fields imply that its
master equation contains the terms describing the process of energy
dissipation from inflaton system to other particles. Barera and Fang
\cite{Berera} initially assumed that consistent description of the
inflaton field with energy dissipation requires the master equation
in form of the Langeven equation, where the fluctuation-dissipation
relation is present, which uniquely links the field fluctuations and the
dissipation of energy. Such equation lies in the the basis of
description of the warm inflation process.

Interaction between the components in the Universe must be
introduced in such a way that preserves the covariance of the
energy-momentum tensor $T^{\,\,\mu\nu}_{(tot)\,\,;\nu}=0,$ therefore
$T^{\,\,\mu\nu}_{_{DM}\,\,;\nu}=-T^{\,\,\mu\nu}_{_{DE}\,\,;\nu}\neq
0,$ where $u_\nu$ is the $4$-velocity. The conservation equations in
that case take on the form:
\begin{equation}\label{ref1}
 u_\nu T^{\,\,\mu\nu}_{_{DM}\,;\mu}
 =-u_\nu T^{\,\,\mu\nu}_{_{DE}\,;\mu}=
 -Q.
\end{equation}
For the FLRW-metric the equations \eqref{ref1} transform to:
\begin{eqnarray}
  &&\dot{\rho}_m+3H\rho_m=Q,\label{eq55}\\
 &&\dot{\rho}_{_{DE}}+3H\rho_{_{DE}}(1+w_{_{DE}})=-Q,\label{eq56}
\end{eqnarray}
where $\rho_m$ and $\rho_{_{DE}}$ are densities of the DM and the
DE respectively, $w_{_{DE}}$ is the EoS parameter for the DE, $H\equiv\dot{a}/a$ is the Hubble
parameter. If $Q < 0,$ then DM continuously decays into DE, and vice versa if $Q> 0$.
Note that the equations \eqref{eq55} and \eqref{eq56} can be combined to obtain the
conservation equation:
\begin{equation}\label{eq57}
     \dot{\rho}_{tot}+3H\rho_{tot}(\rho_{tot}+p_{tot})=0,
\end{equation}
where $\rho_{tot}=\rho_{_{DE}}+\rho_m$ is the total energy density.

The interaction between the DM and the DE is
effectively equivalent to modification of the EoS for the
interacting components. Indeed, the equations \eqref{eq55} can be
transformed to the standard form of the conservation law written for
uncoupled components:
\[\begin{array}{l} {\dot{\rho }_{_{DE}} +3H\rho _{_{DE}} (1+w_{_{DE,eff}} )=0} \\ {\dot{\rho }_{_{DM}} +3H\rho _{_{DM,eff}} =0}, \end{array},\]
where
\begin{equation}\label{w_eff}w _{_{DE,eff}} =w_{_{DE}} -\frac{Q}{3H\rho_{_{DE}} } ;\quad w_{_{DM,eff}} =\frac{Q}{3H\rho _{_{DM}} }, \quad \end{equation}
play role of effective EoS for the DE and the DM respectively.

As we do not know the nature of both the DE and the DM, we cannot derive the coupling $Q$ from the first principles
\cite{secondref}. However it is clear from the dimension analysis that
this quantity must depend on the energy density of one of the dark
component, or a combination of both components with dimension of
energy density, times the quantity of inverse time dimension. For
the latter it is natural to take the Hubble parameter $H.$

The following forms of $Q$ are the most often used in the literature:
 \begin{equation}\label{Q}
    Q_I=3H\gamma \rho_{m}, Q_{II}=3H\gamma \rho_{_{DE}}, Q_{III}=3H\gamma (\rho_{m}+\rho_{_{DE}}).
 \end{equation}
We consider below the simplest case $Q_I$:
\[
\begin{array}{l}
 \dot \rho _{m}  + 3H\rho _{m}  = \delta H\rho _{m},  \\
 \dot \rho _{_{DE}}  + 3H\left( {\rho _{_{DE}}  + p_{_{DE}} } \right) =  - \delta H\rho
 _{m},
 \end{array}
\]
where $\delta $ is the dimensionless coupling constant.

Integration of the latter equations yields
  \begin{equation}\label{Hrho_delta}
   \begin{array}{l}
     \rho _{_{DE}}  = \rho _{_{DE0}} a^{ - \left[ {3(1 + w_{_{DE}} ) + \delta } \right]} ; \hfill  \\
      \rho _{m}  = \frac{{ - \delta \rho _{_{DE0}} }}
{{3w_{_{DE}} + \delta }}a^{ - \left[ {3(1 + w_{_{DE}} ) + \delta }
\right]}  + \left( {\rho _{m0}  + \frac{{\delta \rho _{_{DE0}}
}} {{3w_{_{DE}} + \delta }}} \right)a^{ - 3}.
   \end{array}
  \end{equation}
Inserting the obtained densities into the first Friedmann equation
 and transforming from scale factor to redshift, one
gets for $H^2(z)$ the following
\begin{equation}\label{H2_delta}
H^2 = \frac{(1+z)^3H_0^2}{3(3w_{_{DE}}+\delta)}\left[
3w_{_{DE}}\Omega_{DE}(1+z)^{(3w_{_{DE}}+\delta)}+\Omega_{m}(3w_{_{DE}}+\delta)\right].
\end{equation}
Note that in the considered case the DP $q(z)$
can be easily determined by the following
\[ q(z)=\frac{1+z}{2 H^2}\frac{d H^2}{dz}-1,\] and direct substitution of \eqref{H2_delta} results in the
explicit dependence of the DP on the redshift:
\begin{equation}
   q(z) =-1 +\frac{3}{2}\frac{ w_{_{DE}}\Omega_{DE}(3(1+w_{_{DE}})+\delta)(1+z)^{(3w_{_{DE}}+\delta)}+\Omega_{m}+\delta/w_{_{DE}}}
   {3w_{_{DE}}\Omega_{DE}(1+z)^{(3w_{_{DE}}+\delta)}+\Omega_{m}+\delta/w_{_{DE}}}.
  \end{equation}
With $w_{_{DE}}=-1$ and $\delta = 0,$ this expression coincides with
the DP value obtained in the SCM.

Let us now consider the inverse problem: instead of the interaction
we postulate the ratio
\begin{equation}\label{eq76}
     \frac{\rho_{m}}{\rho_{_{DE}}}=f(a),
\end{equation}
where $f(a)$ is arbitrary differentiable function of the scale factor. The interaction then takes on the form (see \cite{Bolotin_et_al} for details):
\begin{equation}\label{eq71}
Q=H\rho_{_{DE}}\Omega_m\left(\frac{f^\prime a}{f}-3w_{_{DE}}\right)
=H\rho_m\Omega_{_{DE}}\left(\frac{f^\prime a}{f}-3w_{_{DE}}\right).
\end{equation}
Let us define $f(a)$ as
\begin{equation}\label{eq79}
     f(a)=f_0\,a^\xi,
\end{equation}
where $\xi$ is a constant and $f_0$ is the present value of $f(a)$. The
integration constant is \cite{Bolotin_et_al}
\begin{equation}\label{eq80}    \rho_{tot}=\rho_{tot,0}\,a^{-3}\left[\,\Omega_{m0}+
\left(1-\Omega_{m0}\right)a^\xi\,\right]^{-3w_{eff}/\,\xi}.
\end{equation}
Inserting this expression into the Friedmann equation, one finds:
\begin{equation}\label{eq81}
\begin{gathered}
E^2=\frac{H^2}{H_0^2}=a^{-3}\left[\,\Omega_{m0}+\left(1-\Omega_{m0}\right)a^\xi\,
 \right]^{-3w_{_X}/\,\xi}\\
 =(1+z)^3 \left[\,\Omega_{m0}+\left(1-\Omega_{m0}\right)
 (1+z)^{-\xi}\,\right]^{-3w_{_X}/\,\xi}.
\end{gathered}
\end{equation}
Using the above obtained formulae, it is easy to find the following
expression for the DP in the considered model
\begin{equation}\label{q_f_a}
q = 1+\frac{3}{2}\left(w_{_{DE}}\Omega_{_{DE}} +
w_m\Omega_{m}\right),
\end{equation}
where relative densities $\Omega_{_{DE}},~\Omega_{_{DM}}$ have the
following form
\begin{equation}\label{omega_a}
\Omega_{_{DE}}=
\frac{\Omega_{_{DE0}}}{\Omega_{_{DE0}}+(1-\Omega_{_{DE0}})a^{\xi}},~~~\Omega_{_{DM}}=
\frac{(1-\Omega_{_{DE0}})a^{\xi}}{\Omega_{_{DE0}}+(1-\Omega_{_{DE0}})a^{\xi}},
\end{equation}
and we finally obtain
\begin{equation}\label{q_f_a_fin}
q =
\frac{1}{2}+\frac{3}{2}\frac{w_{_{DE}}\Omega_{_{DE0}}(1+z)^{\xi}}{1-\Omega_{_{DE0}}+\Omega_{_{DE0}}(1+z)^{\xi}},
\end{equation}

In the considered model, unlike the SCM, the DE density is not constant, so
it facilitates solution of the coincidence problem
under condition $\xi<3.$ The
accelerated expansion phase, defined by the relation $q(z^*)=0,$ starts at the redshift
\begin{equation}\label{z_f_a_fin}
  z^*=\left(\frac{1-\Omega_{_{DE0}}}{{(1+3w_{_{DE}}})\Omega_{_{DE0}}}\right)^{\frac{1}{\xi}} -1.
\end{equation}
This relation naturally generalizes the corresponding result for the SCM:
\begin{equation}\label{z_trSCM}
z^*={{\left( \frac{2{{\Omega }_{\Lambda 0}}}{{{\Omega }_{m0}}}
\right)}^{1/3}}-1.
\end{equation}
The difference between the values $z^*$ in \eqref{z_f_a_fin} and \eqref{z_trSCM}
depend on how much the parameters $\xi$ and $w_{_{DE}}$ differ from $3$ and $-1$
respectively. In the case of zero difference the quantities
\eqref{z_trSCM} and \eqref{z_f_a_fin} coincide. Note that at $z\to
\infty$ Universe expands with deceleration \[q\to \frac12,\] the case
$z\to -1$ corresponds to \[q\to \frac12-\frac23 w_{_{DE}}.\] Therefore,
as expected, dynamics of the considered model is
asymptotically (at $z\to \infty$ and $z\to -1$) identical to the
case of the two-component Universe, filled by the non-relativistic matter
and the DE with the EoS $p = w_{_{DE}}\rho.$

\end{document}